\documentclass[pdftex,twocolumn,epjc3]{svjour3}          

\RequirePackage[T1]{fontenc}
\RequirePackage[latin1]{inputenc}  
\RequirePackage{xspace}            

\smartqed  

\RequirePackage{graphicx}
\RequirePackage{flushend}
\RequirePackage[numbers,sort&compress]{natbib}
\RequirePackage[colorlinks,citecolor=blue,urlcolor=blue,linkcolor=blue]{hyperref}

\usepackage[symbol]{footmisc}
\usepackage{feynmf}
\usepackage{multirow}
\usepackage{textpos}
\usepackage[switch]{lineno}
\usepackage{afterpage}
\usepackage{amsmath}
\usepackage{siunitx}
\usepackage{amssymb}
  
\journalname{Eur. Phys. J. C}

\def\eeqq{\ensuremath{e^{-}e^{+}\rightarrow q\bar{q}}\xspace}

\begin{document}

\newcommand{\todo}[1]{\textcolor{black}{{#1}}}
\newcommand{\todogreen}[1]{\textcolor{green}{{#1}}}
\newcommand{\todomagenta}[1]{\textcolor{magenta}{{#1}}}

\def\DC{\ensuremath{DC}\xspace}
\def\DT{\ensuremath{DT}\xspace}
\def\DTC{\ensuremath{DTC}\xspace}

\def\Gammaqq{\ensuremath{\Gamma_{q\bar{q}}}\xspace}
\def\Gammahad{\ensuremath{\Gamma_{hadrons}}\xspace}

\def\Kgamma{\ensuremath{E_{\gamma}}\xspace}
\def\Kcut{\ensuremath{E_{\gamma}^{cut}}\xspace}
\def\acolcut{\ensuremath{\sin{(\Psi_{acol})}^{cut}}\xspace}

\def\Kreco{\ensuremath{K_{reco}}\xspace}
\def\kaonness{\ensuremath{\Delta_{\dEdx-K}}\xspace}
\def\pionness{\ensuremath{\Delta_{\dEdx-\pi}}\xspace}
\def\protonness{\ensuremath{\Delta_{\dEdx-p}}\xspace}
\def\epsilonhad{\ensuremath{\epsilon_{had}}\xspace}
\def\epsilonb{\ensuremath{\epsilon_{b}}\xspace}
\def\epsilonc{\ensuremath{\epsilon_{c}}\xspace}
\def\epsilonuds{\ensuremath{\epsilon_{uds}}\xspace}
\def\epsilonb2{\ensuremath{\epsilon^{2}_{b}}\xspace}
\def\epsilonc2{\ensuremath{\epsilon^{2}_{c}}\xspace}
\def\epsilonuds2{\ensuremath{\epsilon^{2}_{uds}}\xspace}

\def\costheta{\ensuremath{\cos \theta}\xspace}
\def\costhetab{\ensuremath{\cos \theta_{b}}\xspace}
\def\costhetaq{\ensuremath{\cos \theta_{q}}\xspace}
\def\sinthetaq{\ensuremath{\sin \theta_{q}}\xspace}
\def\costhetasq{\ensuremath{\cos^2 \theta_{b}}\xspace}
\def\sinthetasq{\ensuremath{\sin^2 \theta_{b}}\xspace}
\def\costhetaj{\ensuremath{\cos \theta_{j}}\xspace}
\def\costhetac{\ensuremath{\cos \theta_{c}}\xspace}

\def\fb{fb\ensuremath{^{-1}}\xspace}
\def\mum{\textmu\ensuremath{m}\xspace}

\def\eL{\ensuremath{e_L^{-}}\xspace}
\def\eR{\ensuremath{e_R^{-}}\xspace}
\def\pL{\ensuremath{e_L^{+}}\xspace}
\def\pR{\ensuremath{e_R^{+}}\xspace}

\def\b{\ensuremath{b}\xspace}
\def\bbar{\ensuremath{\overline{b}}\xspace}
\def\bbbar{\ensuremath{b}\ensuremath{\overline{b}}\xspace}
\def\qqbar{\ensuremath{q}\ensuremath{\overline{q}}\xspace}
\def\ccbar{\ensuremath{c}\ensuremath{\overline{c}}\xspace}
\def\ttbar{\ensuremath{t}\ensuremath{\overline{t}}\xspace}
\def\eebbbar{\ensuremath{e^{-}e^{+}\rightarrow b\bar{b}}\xspace}
\def\eeccbar{\ensuremath{e^{-}e^{+}\rightarrow c\bar{c}}\xspace}
\def\eeqqbar{\ensuremath{e^{-}e^{+}\rightarrow q\bar{q}}\xspace}
\def\ee{\ensuremath{e^{-}e^{+}}\xspace}
\def\eeZ{\ensuremath{e^{-}e^{+}\rightarrow Z}\xspace}
\def\eeZqqbar{\ensuremath{e^{-}e^{+}\rightarrow Z\rightarrow q\bar{q}}\xspace}
\def\eeZgammaqqbar{\ensuremath{e^{-}e^{+}\rightarrow Z \gamma \rightarrow q\bar{q} \gamma}\xspace}

\def\eebb{\ensuremath{e^{-}e^{+}\rightarrow b\bar{b}}\xspace}
\def\eecc{\ensuremath{e^{-}e^{+}\rightarrow c\bar{c}}\xspace}
\def\eeZqq{\ensuremath{e^{-}e^{+}\rightarrow Z\rightarrow q\bar{q}}\xspace}
\def\eeZgammaqq{\ensuremath{e^{-}e^{+}\rightarrow Z \gamma \rightarrow q\bar{q} \gamma}\xspace}

\def\cme{\ensuremath{\sqrt{s}}\xspace}
\def\eLpR{\ensuremath{e_L^{-}e_R^{+}}\xspace}
\def\eRpL{\ensuremath{e_R^{-}e_L^{+}}\xspace}
\def\eLpRqq{\ensuremath{e_L^{-}e_R^{+}\rightarrow q\bar{q}}\xspace}
\def\eRpLqq{\ensuremath{e_R^{-}e_L^{+}}\rightarrow q\bar{q}\xspace}

\def\dEdx{\ensuremath{dE/\/dx}\xspace}
\def\dNdx{\ensuremath{dN/\/dx}\xspace}
\def\Afbb{\ensuremath{A^{b}_{FB}}\xspace}
\def\AFBb{\ensuremath{A^{b}_{FB}}\xspace}
\def\Afb{\ensuremath{A_{FB}}\xspace}
\def\AFB{\ensuremath{A_{FB}}\xspace}

\def\ALR{\ensuremath{A_{LR}}\xspace}
\def\Rb{\ensuremath{R_{b}}\xspace}
\def\Rc{\ensuremath{R_{c}}\xspace}
\def\Rq{\ensuremath{R_{q}}\xspace}
\def\Rqp{\ensuremath{R_{q\prime}}\xspace}
\def\Ruds{\ensuremath{R_{uds}}\xspace}
\def\Rbcostheta{\ensuremath{R_{b}(|cos\theta_{b}|)}\xspace}
\def\Rccostheta{\ensuremath{R_{c}(|cos\theta_{c}|)}\xspace}
\def\Rqcostheta{\ensuremath{R_{q}(|cos\theta_{q}|)}\xspace}
\def\Rudscostheta{\ensuremath{R_{uds}(|cos\theta_{uds}|)}\xspace}
\def\Afbc{\ensuremath{A^{c}_{FB}}\xspace}
\def\Afbq{\ensuremath{A^{q}_{FB}}\xspace}
\def\AFBc{\ensuremath{A^{c}_{FB}}\xspace}
\def\AFBq{\ensuremath{A^{q}_{FB}}\xspace}
\def\eett{\ensuremath{e^{\mbox{\scriptsize -}}}\ensuremath{e^{\mbox{\scriptsize +}}}\ensuremath{\rightarrow}\ensuremath{t}\ensuremath{\overline{t}}\xspace}
\def\bquark{\ensuremath{b}-quark\xspace}
\def\bjet{\ensuremath{b}-jet\xspace}
\def\bjets{\ensuremath{b}-jets\xspace}
\def\btagging{\ensuremath{b}-tagging\xspace}
\def\btag{\ensuremath{b_{tag}}\xspace}
\def\cquark{\ensuremath{c}-quark\xspace}
\def\cjet{\ensuremath{c}-jet\xspace}
\def\cjets{\ensuremath{c}-jets\xspace}
\def\ctagging{\ensuremath{c}-tagging\xspace}
\def\ctag{\ensuremath{c_{tag}}\xspace}
\def\udsjet{\ensuremath{uds}-jet\xspace}
\def\udsjets{\ensuremath{uds}-jets\xspace}
\def\tquark{\ensuremath{t}-quark\xspace}
\def\Zboson{\ensuremath{Z}-boson\xspace}
\def\Zpole{\ensuremath{Z}-pole\xspace}
\def\Zprime{\ensuremath{Z^{\prime}}\xspace}
\def\Zbb{\ensuremath{Z_{b\bar{b}}}\xspace}
\def\ZbRbR{\ensuremath{Z_{b_{R}\bar{b}_R}}\xspace}
\def\ZbLbL{\ensuremath{Z_{b_{L}\bar{b}_L}}\xspace}
\def\Bc{\ensuremath{Vtx}-method\xspace}
\def\Kc{\ensuremath{K}-method\xspace}
\def\BcKc{\ensuremath{Vtx/K}-method\xspace}
\def\BcKcsame{\ensuremath{Vtx/K_{same\,jet}}-method\xspace}
\def\BcBc{\ensuremath{Vtx/Vtx}-method\xspace}
\def\KcKc{\ensuremath{K/K}-method\xspace}

\def\Pb{\ensuremath{P_{chg}}\xspace}
\def\Pbb{\ensuremath{P^b_{chg}}\xspace}
\def\Pbc{\ensuremath{P^c_{chg}}\xspace}
\def\Qb{\ensuremath{Q_{chg}}\xspace}

\def\PbB{\ensuremath{P_{chg,M_{1}}}\xspace}
\def\PbK{\ensuremath{P_{chg,M_{2}}}\xspace}
\def\QbB{\ensuremath{Q_{chg,M_{1}}}\xspace}
\def\QbK{\ensuremath{Q_{chg,M_{2}}}\xspace}

\def\Pbi{\ensuremath{P_{chg,i}}\xspace}
\def\Pbj{\ensuremath{P_{chg,j}}\xspace}
\def\Qbi{\ensuremath{Q_{chg,i}}\xspace}
\def\Qbj{\ensuremath{Q_{chg,j}}\xspace}

\def\pmp{\ensuremath{+-}\xspace}
\def\mpp{\ensuremath{-+}\xspace}
\def\pp{\ensuremath{++}\xspace}
\def\mm{\ensuremath{--}\xspace}

\def\Aone{\ensuremath{A_{1}}\xspace}
\def\Atwo{\ensuremath{A_{2}}\xspace}
\def\Athree{\ensuremath{A_{3}}\xspace}
\def\Amodel{\ensuremath{A} model\xspace}
\def\Amodels{\ensuremath{A} models\xspace}
\def\Bmodel{\ensuremath{B} model\xspace}
\def\Bmodels{\ensuremath{B} models\xspace}
\def\PlusMinusmodel{\ensuremath{X^{\pm}} model\xspace}
\def\PlusMinusModel{\ensuremath{X^{\pm}} model\xspace}
\def\PlusMinusmodels{\ensuremath{X^{\pm}} models\xspace}
\def\PlusMinusModels{\ensuremath{X^{\pm}} models\xspace}
\newcommand{\NM}[2]{\ensuremath{{#1}^{{#2}}}\xspace}
\def\ILCnopol{ILC250\ensuremath{^\blacklozenge(no~pol.)}\xspace}

\def\Bm{\ensuremath{B}\xspace}
\def\BL{\ensuremath{B_{L}}\xspace}
\def\BH{\ensuremath{B_{H}}\xspace}
\def\Bplus{\ensuremath{B_{+}}\xspace}
\def\Bminus{\ensuremath{B_{-}}\xspace}

\newcommand\blankpage{%
    \null
    \thispagestyle{empty}%
    \addtocounter{page}{-1}%
    \newpage}

\title{Probing Gauge-Higgs Unification models at the ILC with quark-antiquark forward-backward asymmetry at center-of-mass energies above the $Z$ mass.
\thanksref{t1}}


\author{A. Irles\thanksref{addr1,e1}
        \and J.P. M\'arquez\thanksref{addr1} 
        \and R. P\"oschl\thanksref{addr2}
        \and F. Richard\thanksref{addr2}
        \and A. Saibel\thanksref{addr1} \and
        \\ H. Yamamoto\thanksref{e3,addr1} 
        \and N. Yamatsu\thanksref{addr3,addr4}
}

\thankstext[$\star$]{t1}{This work was carried out in the framework of the ILD concept group.}
\thankstext{e1}{e-mail: adrian.irles@ific.uv.es (corresponding author)}
\thankstext{e3}{On leave from Tohoku University, Sendai, Japan}

\institute{IFIC, Universitat de Val\`encia and CSIC, C./ Catedr\'atico Jos\'e Beltr\'an 2, E-46980 Paterna, Spain\label{addr1}
\and Universit{\'e} Paris-Saclay, CNRS/IN2P3, IJCLab, 91405 Orsay, France\label{addr2}
\and Department of Physics, National Taiwan University, Taipei, Taiwan 10617, R.O.C.\label{addr3}
\and Yukawa Institute for Theoretical Physics, Kyoto University, Kitashirakawa Oiwakecho, Sakyo-ku, Kyoto 606-8502, Japan.\label{addr4}
}

\date{Received: date / Accepted: date}

\maketitle

\begin{abstract}
The International Linear Collider (ILC) will allow the precise study of $\ee\rightarrow q\bar{q}$ interactions at different center-of-mass energies from the \Zpole\ to 1 TeV. 
In this paper, we discuss the experimental prospects for measuring differential observables in $\ee\rightarrow b\bar{b}$ and $\ee\rightarrow c\bar{c}$ at the ILC baseline energies, 250 and 500 GeV. 
The study is based on full simulation and reconstruction of the International Large Detector (ILD) concept. 
Two gauge-Higgs unification models predicting new high-mass resonances beyond the Standard Model are discussed.
These models predict sizable deviations of the forward-backward observables at the ILC running above the $Z$ mass and with longitudinally polarized electron and positron beams.
The ability of the ILC to probe these models via high-precision measurements of the forward-backward asymmetry is discussed.
Alternative scenarios at other energies and beam polarization schemes are also discussed, extrapolating the estimated uncertainties from the two baseline scenarios.

\keywords{ILC \and Beyond Standard Model \and Higgs-Boson}
\end{abstract}

\section{Introduction}
\label{sec:intro}

The Standard Model (SM) is a successful theory, well-established experimentally and theoretically. With the discovery of the Higgs boson \cite{CMS:2012qbp,ATLAS:2012yve}, the structure of the SM seems to be confirmed. 
However, the SM cannot explain many of its seemingly arbitrary features.
An example is the striking mass hierarchy in the fermion sector. 
Moreover, while the dynamics of the SM gauge bosons, the photon, $W$ and $Z$ bosons, and gluons are governed by the gauge principle, the dynamics of the Higgs boson are different and unique in the SM. 
The SM does not predict the strength of the Higgs couplings of quarks and leptons, nor the Higgs self-couplings. 
Large quantum corrections must be canceled by fine-tuning the parameters to match the measured Higgs boson mass. 
One possible solution to this issue, achieving stabilization of the Higgs mass against quantum corrections, appears when the Higgs boson is associated with the zeroth mode of a dimension-five component of extensions of the SM gauge group. 
These models are referred to as gauge-Higgs unification (GHU) models.

The two most precise determinations of $\sin^{2}\theta_{\text{eff}}$ by the LEP and SLC differ by $3.7$ standard deviations, and neither agrees with the SM prediction \cite{ALEPH:2005ab,Djouadi:2006rk}. In particular, the LEP value was extracted from the forward-backward asymmetry measurement for $b$-quarks in LEP1 data, and is nearly three standard deviations away from the value predicted by the SM. 
Clarifying this anomaly and exploring the possibility of BSM physics motivates the study of quark pair production in high energy $\ee$ collisions at future colliders both at the $Z$ boson mass and higher energies.
In the SM, these interactions are mediated by the photon, $Z$ boson, and their interference. 
Some BSM theories predict deviations of these bosons' couplings or even sizable new contributions to these processes from new mediators (such as heavy $Z^{\prime}$ resonances).
These deviations would be accessible experimentally by performing high precision measurements of $\ee\rightarrow\qqbar$ observables at different center-of-mass energies (\cme). 
The work presented here is based on the study of such processes at the ILC.

In parallel to the exploitation of data from the Large Hadron Collider (LHC), the high-energy accelerator-based particle physics community is working towards the next large colliders after the LHC. \todo{On one hand, there is the Electron-Ion Collider (EIC) which has been recently approved for construction at Brookhaven National Laboratory \cite{Accardi:2012qut}. On the other hand, various projects of high-energy $\ee$ colliders have been proposed and are under discussion.}
These ``Higgs Factories'' are designed for the precise scrutiny of the Higgs sector and search for new physics through precision measurements. 
We here discuss in detail the International Linear Collider (ILC) which we consider the most mature from a technological point of view, having produced a technical design report (TDR) in 2013 \cite{Behnke:2013xla,Baer:2013cma,Adolphsen:2013jya, Adolphsen:2013kya,Behnke:2013lya}. 
The ILC plans for a comprehensive high-precision physics program based on collisions of polarized electron and positron beams at a center-of-mass energy of 250 ``ILC250'' and 500 GeV ``ILC500''.
Operation around the \ttbar\ production threshold, the $Z$-mass, and at 1 TeV are also proposed. 
The studies discussed in this document are based on full simulations of the International Large Detector (ILD) concept \cite{Behnke:2013lya,ILD:2020qve}. 
The ILD is one of the detectors proposed for collecting and exploiting the ILC data. It has been optimized to perform high-precision measurements at ILC250 and ILC500. 

Other proposals for Higgs factories, not discussed in this document, are the Compact Linear Collider (CLIC \cite{CLIC:2016zwp,CLIC:2018fvx,CLICdp:2018cto}), the Future Circular \ee\ Collider (FCCee \cite{FCC:2018byv,FCC:2018evy}), the Circular electron-positron Collider (CEPC \cite{CEPCStudyGroup:2018rmc,CEPCStudyGroup:2018ghi}) and the Cool Copper Collider (C$^{3}$ \cite{Vernieri:2022fae}). 
The CLIC and C$^{3}$ are linear colliders with baseline design collision energies of 380, 1000, and 3000 GeV (for CLIC) or 250 and 550 GeV (for  C$^{3}$). 
Both foresee using longitudinal polarization only for the electron beam. 
The FCCee and CEPC are electron-positron circular colliders featuring only longitudinally unpolarized beams and with baseline collision energies at the $Z$ mass and 240 GeV with an eventual upgrade to 365 GeV, at least for the FCCee case. Intermediate energy stages around the $WW$ production threshold are also envisioned.

The content of this article is structured as follows.
Section \ref{sec:theory} briefly describes the two GHU models used as benchmarks and discusses the sensitivity of the forward-backward asymmetry, \AFB, in \eebb and \eecc processes at different energies and beam polarization scenarios. 
Section \ref{sec:exp_framework} describes the ILC, the ILD, and the experimental framework for the study presented here.
Section \ref{sec:AFB} presents a detailed discussion of the methodology and prospects for precisely measuring \AFBc and \AFBb at ILC250 and ILC500, based on previous full simulation studies~\cite{Irles:2023ojs,Irles:2023nee}. 
Section \ref{sec:analysis} discusses the potential of indirect constrains on the two aforementioned GHU theories through the measurement of \AFB. Conclusions and prospects are discussed in Section \ref{sec:summary}.

\section{Theory}
\label{sec:theory}

\subsection{Gauge-Higgs Unification Models}

The $SU(3)_c\times SO(5)_W\times U(1)_X$ GHU models on a
Randall-Sundrum (RS) warped space have been studied as
candidates to explain physics beyond the SM, where $SO(5)_W\times U(1)_X$ contains the
electroweak gauge symmetry $SU(2)_L\times U(1)_Y$~\cite{Agashe:2004rs,Medina:2007hz,Hosotani:2008tx,Funatsu:2014fda,Funatsu:2019xwr,Funatsu:2020haj,Funatsu:2021yrh,Funatsu:2023jng,Yamatsu:2023bde}.
In this scenario, the Higgs field naturally appears as a fluctuation mode of the Aharonov-Bohm phase $\theta_{H}$ in the extra dimension.
Gauge symmetry stabilizes the Higgs boson mass against quantum
corrections.
With suitable parameter choices, low energy predictions of this gauge theory can match the SM.
In this paper, two specific GHU models are considered: the $A$~\cite{Funatsu:2014fda} and
$B$~\cite{Funatsu:2019xwr} models.
\todo{The only GHU models known so far that are consistent with the current
experimental results are the $A$ and $B$ models based on the
$SU(3)_c\times SO(5)_W\times U(1)_X$ symmetry.}

Both models predict massive neutral vector bosons $Z^{\prime}$, which are the Kaluza-Klein (KK) modes of $\gamma$, $Z$, and $Z_{R}$ (the $SU(2)_{R}$ gauge boson).
Here, we simply describe the models and parameter sets that are used in this paper. 
For the full description see Refs.~\cite{Agashe:2004rs,Medina:2007hz,Hosotani:2008tx,Funatsu:2014fda,Funatsu:2019xwr,Funatsu:2020haj,Funatsu:2021yrh,Funatsu:2023jng,Yamatsu:2023bde}. 

\begin{figure*}[!ht]
\centering
    \begin{tabular}{ccc}
            \includegraphics[width=0.3\textwidth]{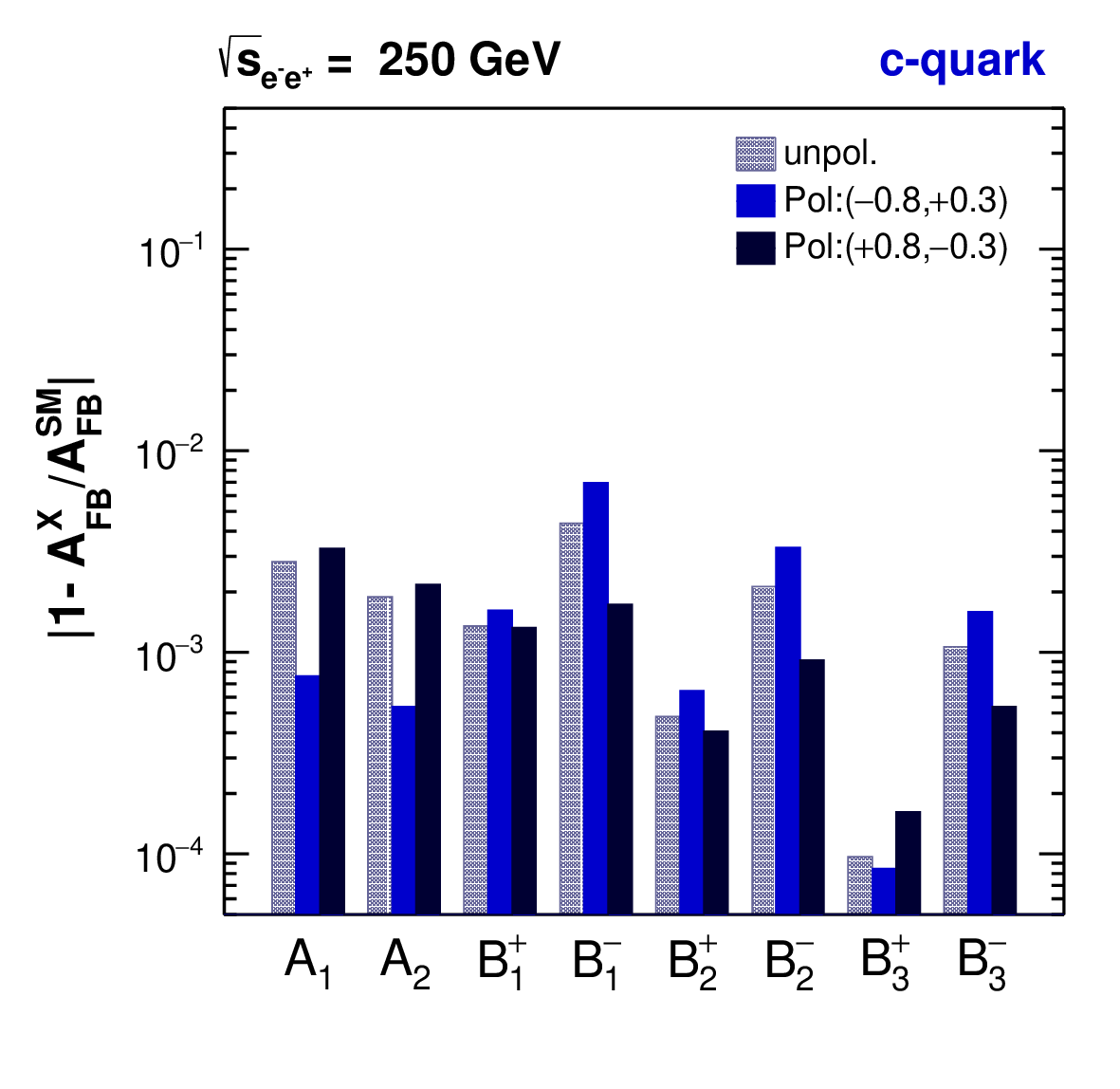} &
            \includegraphics[width=0.3\textwidth]{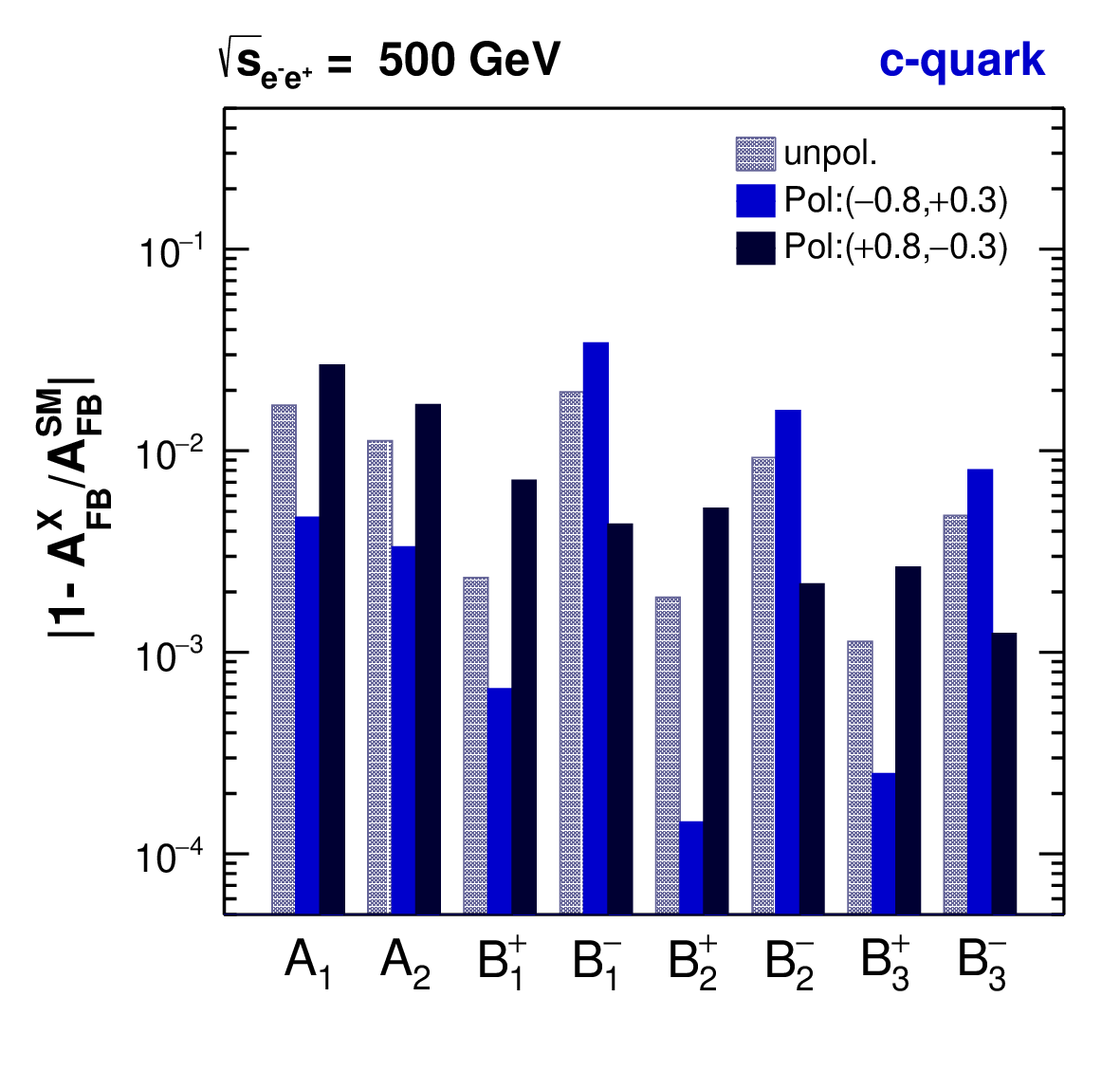} &
            \includegraphics[width=0.3\textwidth]{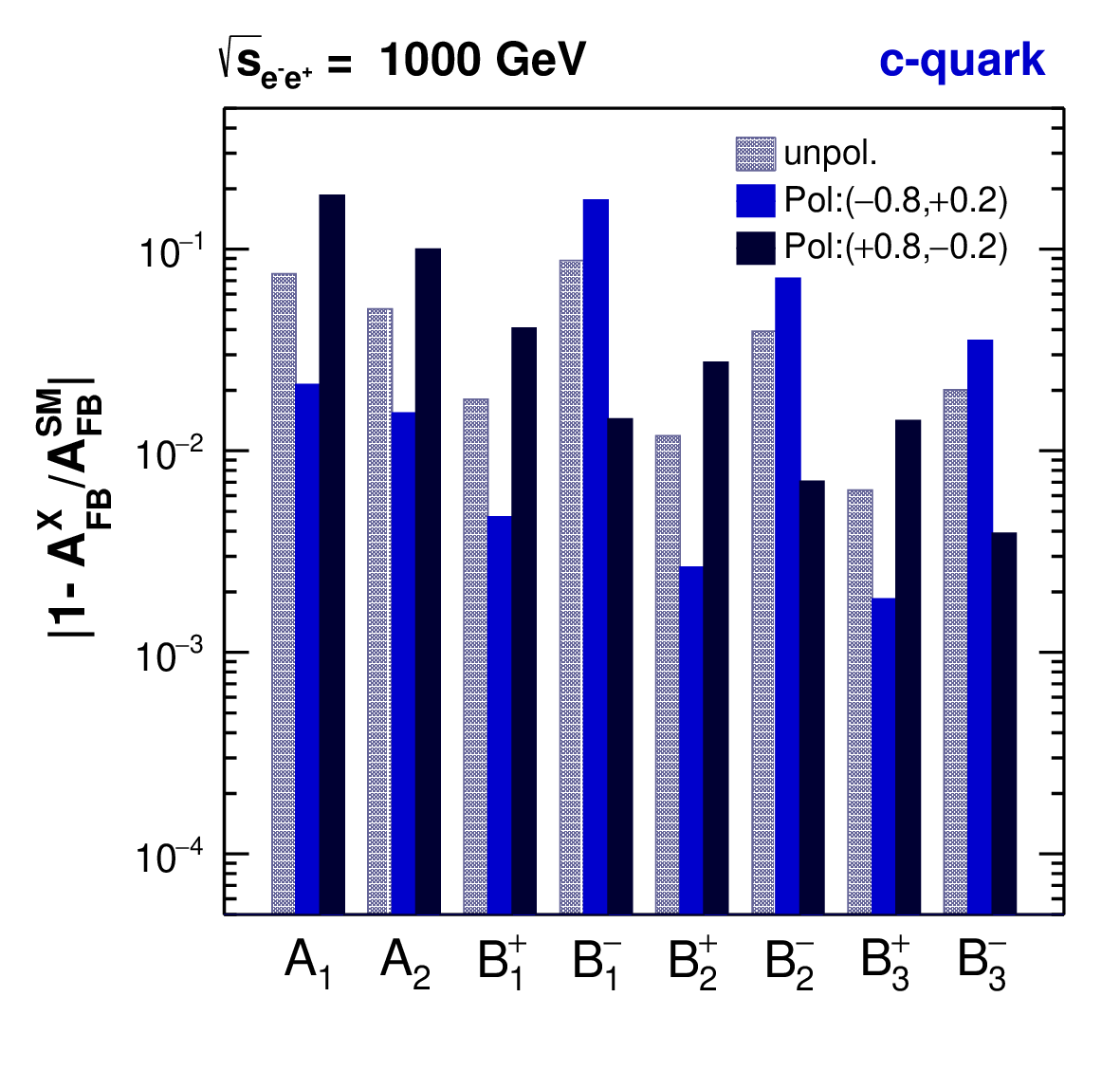} \\ 
            \includegraphics[width=0.3\textwidth]{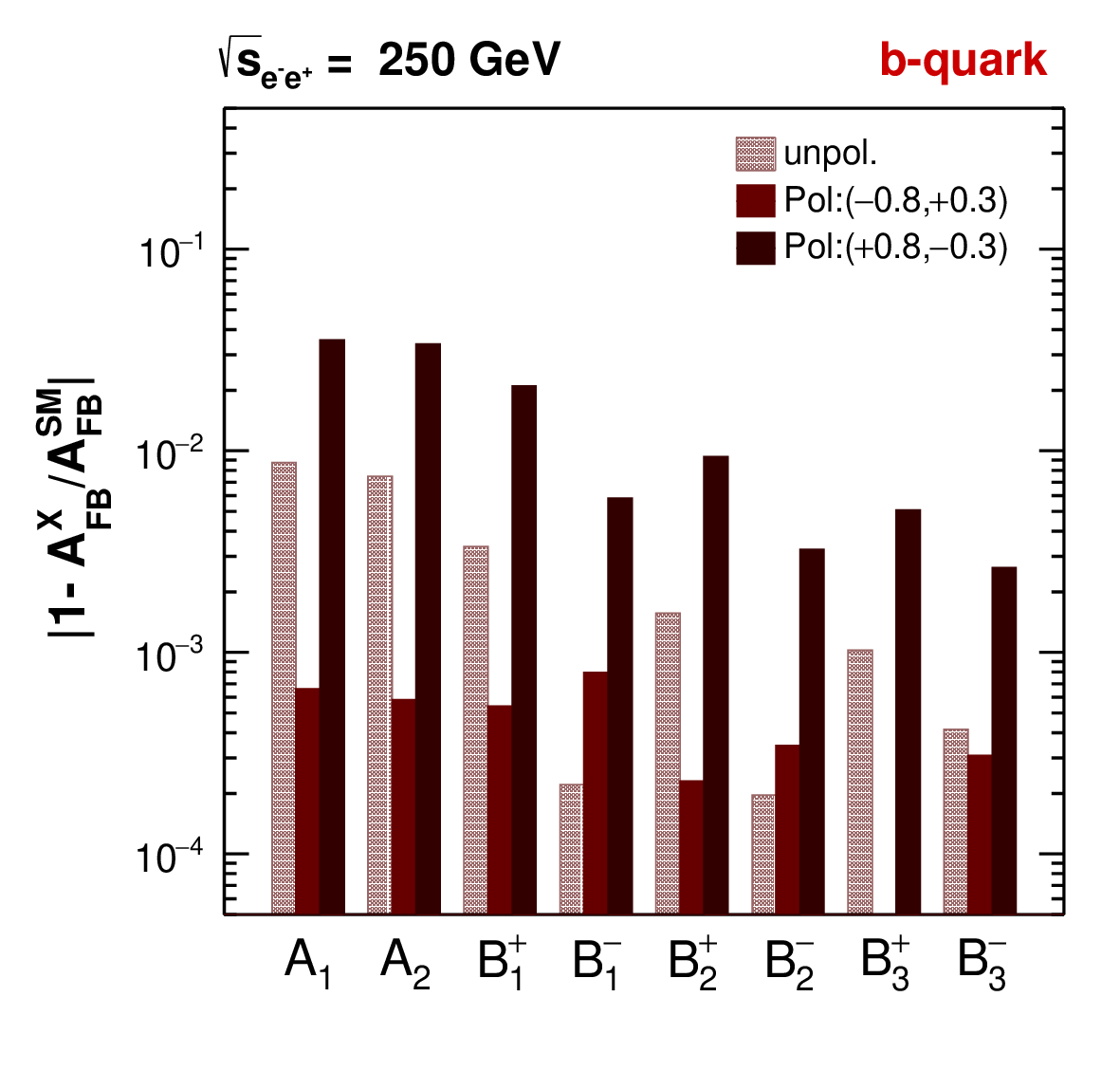} &
            \includegraphics[width=0.3\textwidth]{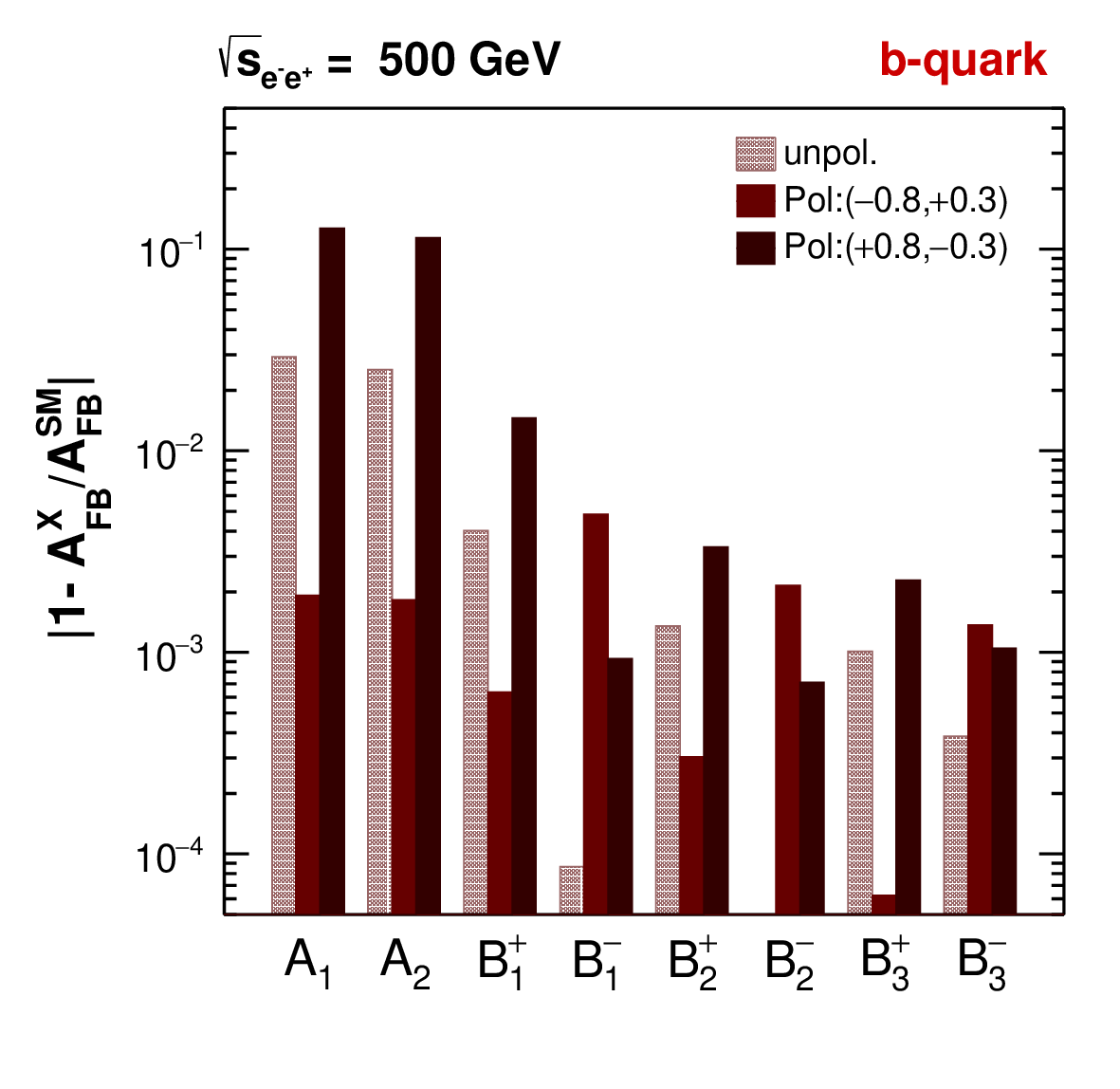} &
            \includegraphics[width=0.3\textwidth]{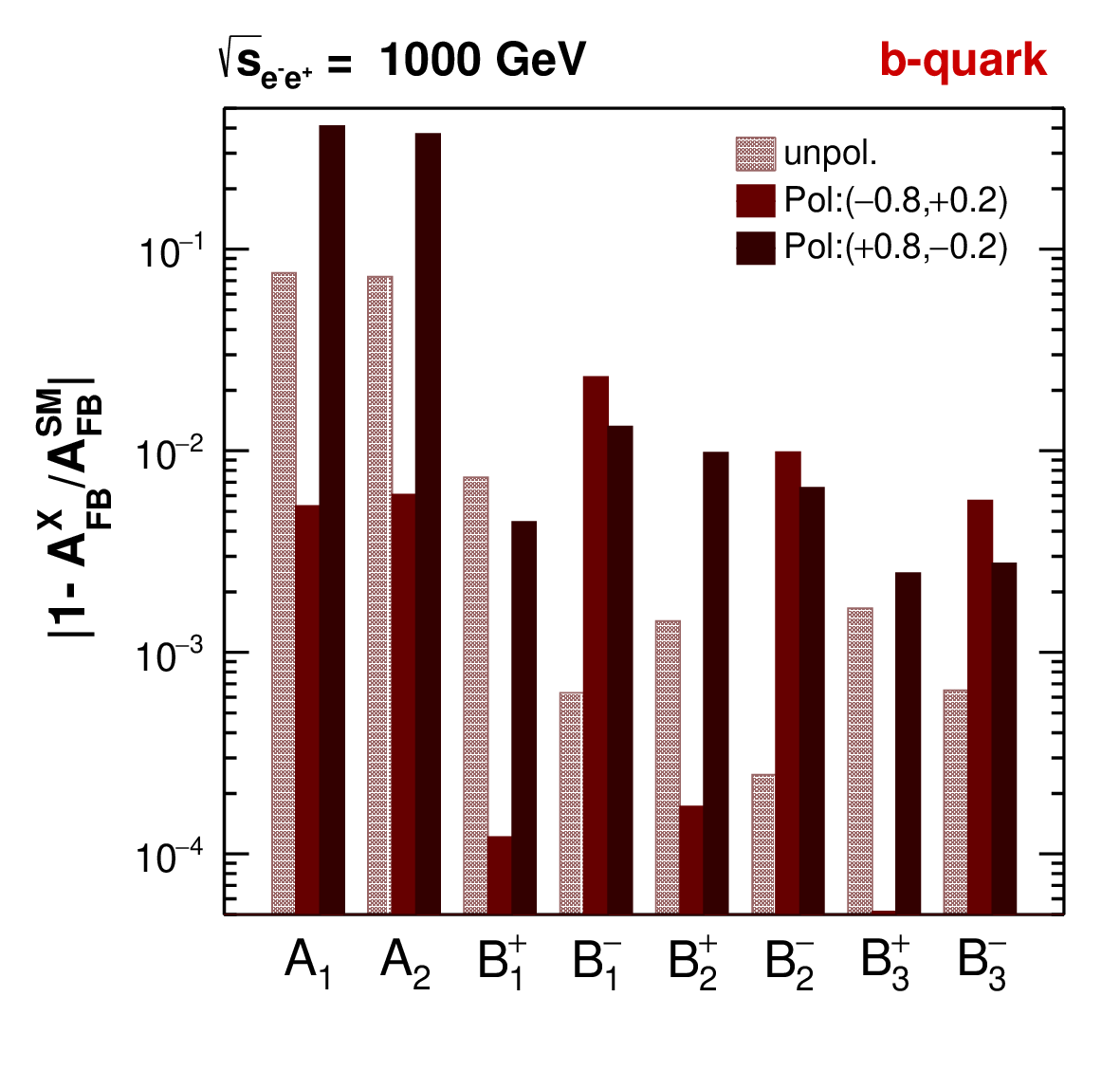} 
    \end{tabular}
    \caption{Predicted differences for the \AFB observable in \ee collisions at several \cme between different GHU models and the SM. 
    The expectations for different final states for \cquark or \bquark pair production and different longitudinal beam polarizations are compared for each energy. 
    }
    \label{fig:theory1}
\end{figure*}

In the $A$ model~\cite{Funatsu:2014fda}, quark-lepton
multiplets are introduced in the vector representation of $SO(5)$. 
The non-observation of $Z^{\prime}$ at the LHC implies limits
$\theta_{H}\lesssim 0.09$, $m_{KK}\gtrsim 9$ TeV~\cite{Funatsu:2021yrh}.
Whether the coupling of the $Z^\prime$ bosons is stronger to right- or left-handed
fermions depends strongly on the sign of the bulk mass of fermions. 
The couplings of the right-handed fermions to the $Z^{\prime}$ are large 
since positive bulk masses of fermions are chosen to avoid large
deviations of the $Z$ boson couplings to fermions.
In this paper, two parameter sets of the $A$ model (\Aone
and \Atwo) are adopted as benchmark points~\cite{Funatsu:2017nfm}: 
\footnote{The parameter sets $\Aone$ and $\Atwo$ correspond to $\Atwo$
and $\Athree$ in~\cite{Funatsu:2017nfm}, respectively.}
\newline
$\Aone:\theta_{H}=0.0917,m_{KK}=8.81$ TeV$\rightarrow m_{Z^{1}}=7.19$ TeV;\newline
$\Atwo:\theta_{H}=0.0737,m_{KK}=10.3$ TeV$\rightarrow m_{Z^{1}}=8.52$ TeV,\newline
where $Z^{1}$ is the first KK $Z$ boson. The masses of the first KK 
$\gamma$ and $Z_R$ bosons are similar to those of the first
KK $Z$ boson.
The parameters are chosen such that the couplings of the SM-like $Z$ boson to
fermions (other than the top quark) agree with current measurements within
one part in $10^4$.
The values of $m_{KK}$ and $\theta_H$ for which the top quark and the Higgs
boson masses and electroweak symmetry breaking can be realized are
strongly restricted. 
The $m_{KK}$ value in the parameter set $A_2$ is
at the maximum reachable in the \Amodel.


In the $B$ model~\cite{Funatsu:2019xwr}, which is inspired by grand unification,
the quark-lepton multiplets are introduced in spinor, vector, and
singlet representations of $SO(5)$. 
More specifically, this model is constructed as a low-energy effective
description of the $SO(11)$ gauge-Higgs grand unification model on
the RS warped space \cite{Hosotani:2015hoa,Furui:2016owe},
and the representations of $SU(3)_c$, $SO(5)_W$, and $U(1)_X$ that can
be introduced are strongly restricted by $SO(11)$ gauge symmetry.
The non-observation of $Z^{\prime}$ and $W^{\prime}$ signals at LHC implies limits $\theta_{H}\lesssim 0.10$, $m_{KK}\gtrsim 13$ TeV \cite{Funatsu:2021yrh}.
\todo{At the High-Luminosity LHC with an integrated luminosity of 3000 $fb^{-1}$ and $\sqrt{s}=14$ TeV, it may be possible to search up to $m_{ KK}\simeq 22$ TeV for the GHU $B$ model~\cite{Funatsu:2021yrh}. However, in the studies of Ref.~\cite{Funatsu:2021yrh}, background processes are ignored, and the acceptances and efficiency are not taken into account.}
In this paper, six parameter sets of the $B$ model, $B_j^\pm$
($j=1,2,3$), are adopted as benchmark points 
\cite{Funatsu:2023jng,Yamatsu:2023bde}
\footnote{The parameter sets $B_1^\pm$, $B_2^\pm$, and $B_3^\pm$ 
correspond to $A^\pm$, $B^\pm$, and $C^\pm$ in
\cite{Funatsu:2023jng,Yamatsu:2023bde}, respectively.}
:\newline
$B_1^\pm$: $\theta_{H}=0.10,m_{KK}=13$ TeV$\rightarrow m_{Z^{1}}=10.2$ TeV;\newline
$B_2^\pm$: $\theta_{H}=0.07,m_{KK}=19$ TeV$\rightarrow m_{Z^{1}}=14.9$ TeV;\newline
$B_3^\pm$: $\theta_{H}=0.05,m_{KK}=25$ TeV$\rightarrow m_{Z^{1}}=19.6$ TeV;\newline
where the superscripts $\pm$ indicate the sign of the lepton bulk
masses. The bulk masses of the quarks are negative in all
parameter sets since very light vector-like quarks appear when the bulk
masses of the quarks are taken to be positive.
Model parameters are chosen such that $Z$ couplings to
fermions (other than the top quark) agree with the SM within one part in $10^3$.
These models make use of the so-called 4D bare Weinberg angle (a
projection of the electroweak mixing angle), which is defined such that
all the parameter sets in the \Bmodel predict values of \AFB$(e^-e^+\rightarrow \mu^-\mu^+)$ compatible with current best measurements~\cite{Funatsu:2020haj}.
For $B_j^+$,  the coupling of right-handed electron to the $Z'$ boson is
larger than that of the left-handed electron; for $B_j^-$,
the coupling of left-handed electron to $Z'$ boson is
larger than that of the right-handed electron.

\subsection{Forward-backward asymmetry predictions in GHU}


The forward-backward asymmetry, \AFB, is defined as

\begin{equation}
A^{q}_{FB}=\frac{\sigma^{F}-\sigma^{B}}{\sigma^{F}+\sigma^{B}},
\label{formula:AFB}
\end{equation}
where $\sigma^{F/B}$ is the $\ee\rightarrow q\bar{q}$ cross-section in the forward (F) and backward (B) hemisphere as defined by the polar angle of the quark $\theta_q$ in the nominal center-of-mass reference frame and with respect to the electron beam direction.
For each of the models described above, \AFB has been calculated at leading order for $b$ and \cquark production in several \ee scenarios at various center-of-mass energies and beam polarizations.
We use the $(P_{e^{-}},P_{e^{+}})$ notation for beam polarization, in which the first term is for the electron and the second for the positron beam, a negative sign signifies a left-handed polarization, and 0 corresponds to un-polarized and $\pm$1 to fully polarized beams.

The deviations from the SM value of \AFB induced by the different models are shown in Fig.~\ref{fig:theory1} at 250, 500, and 1000 GeV, with and without ILC-like beam polarization.
The expected differences increase with energy and show large variations depending on the model and the beam polarization. At 250 GeV, the largest deviations occur in \bquark pair production with $(+0.8,-0.3)$ beam polarization, with the highest values for the \Amodels.
At 500 GeV, \cquark pair production also shows large deviations in $(+0.8,-0.3)$ for the \Amodels and in $(-0.8,+0.3)$ for the \Bmodels.
At 1 TeV, most models show sizeable deviations for
at least one of the discussed channels.


\section{Experimental framework}
\label{sec:exp_framework}

\subsection{The International Linear Collider and the International Large Detector}
\label{sec:ILCILD}

The International Linear Collider (ILC) is a linear electron-positron collider that will produce collisions at several energies, and feature a high degree of longitudinal polarization for both beams.
This article focuses on collisions at center-of-mass energies of 250 GeV and 500 GeV (ILC250 and ILC500) in the baseline running scenario, so-called H20-staged~\cite{Bambade:2019fyw}. 
The H20-staged scenario assumes different integrated luminosities split between left-handed and/or right-handed electron and positron beams. 
In addition, we also briefly discuss other scenarios such as operation at the \Zpole (ILCGigaZ) and 1 TeV (ILC1000). 
This information is summarized in the Tab.~\ref{tab:lumNEW}.




\begin{table}[!ht]
\caption{Considered integrated luminosities,$\int \mathcal{L}$, and beam polarization degree scenarios considered in this work.
The second row gives the degree of beam polarization for electrons and positrons. The third row shows the split of the total integrated luminosities when operating with opposite sign polarization (OSP) or same sign polarization (SSP) beams.\label{tab:lumNEW}}
\begin{tabular}{c|c|c|c|c}
                                                           & ILCGigaZ & ILC250 & ILC500 & ILC1000 \\ \hline
$\int \mathcal{L}$ $[\fb]$ & 100      & 2000   & 4000   & 8000    \\ \hline
$(|P_{e^{-}}|,|P_{e^{+}}|)$             & (0.8,0.3)    & (0.8,0.3) & (0.8,0.3)  & (0.8,0.2)   \\ \hline
OSP|SSP [$\%$]                      & 40|10   & 45|5   & 40|10  & 40|10   \\ \hline
\end{tabular}
\end{table}

The International Large Detector (ILD) is one of the detectors proposed for collecting and exploiting the ILC data. 
The ILD design is optimized for the reconstruction of final state particles using Particle Flow techniques~\cite{Brient:2002gh,Thomson:2009rp}. 
ILD consists of inner vertexing and tracking systems and high granularity calorimeters within a 3.5~T solenoid, followed by an instrumented flux return used to identify muons. 
A detailed description of the different subsystems and the proposed technological solutions can be found in Refs.~\cite{Behnke:2013lya,ILD:2020qve}. 
The tracking systems in the current ILD design are briefly discussed due to their crucial role in the studies presented in this paper.

The vertexing and tracking systems are based on silicon sensors and a time projection chamber (TPC).
The vertex detector (VTX) is the closest to the beam pipe, spanning radii from 16 to 60 mm. 
Its design is optimized to provide a single hit resolution of 3 \textmu m. 
The ILC bunch train structure allows for power-pulsed operation, reducing power consumption and cooling requirements by
one to two orders of magnitude. 
Low-mass passive cooling technologies can therefore be used, resulting in a material budget of around 0.15\% of a radiation length per layer, thereby minimizing multiple scattering.

Silicon tracking systems follow the VTX detector: the silicon internal tracker (SIT) covers the central region, and the forward tracking detector (FTD) extends the coverage to lower angles closer to the beam axis.
The SIT also features a barrel geometry and covers the region between 16 and 164 degrees with respect to the beam axis. 
The FTD comprises disks perpendicular to the beam axis and is designed to cover the low-angle region down to 4.8 degrees, complementing the SIT coverage between 16 and 32 degrees. 
The TPC is a large volume time projection chamber allowing continuous 3D tracking and charged particle identification based on the specific energy loss \dEdx. 
It has a length of 4 m and spans radii from 329 mm to 1808 mm, providing up to 220 track measurements with a position resolution in the $r-\phi$ plane of around 100 \textmu m and a \dEdx resolution of approximately 4.5$\%$ with pad-based readout.
An alternative approach read out by 55~\textmu m pixels has the potential for improved performance.
Simulations extrapolating beam test results show that an improved relative resolution of $\sim3$-$4\%$ will be feasible using cluster counting techniques (\dNdx) instead of the traditional \dEdx approach~\cite{LCTPC:2022pvp}.

\subsection{Event simulation}
\label{sec:simulation}

The results in this paper are based on full simulation samples of ILC250 and ILC500 provided by the ILD concept group. 
All simulations use the ILD-L~\cite{ILD:2020qve} model, whose geometry, material, and readout are implemented in \texttt{DD4HEP}~\cite{Frank:2014zya}, interfaced with \texttt{Geant4}~\cite{Agostinelli:2002hh,Allison:2006ve,Allison:2016lfl}.
Inactive materials describing cables, support structures, and services are accounted for in the simulation and reconstruction. 
SM signal and background events are generated with the \texttt{WHIZARD} v2.8.5~\cite{Kilian:2007gr} event generator at LO, including QED ISR in perturbative leading-logarithmic approximation. 
Parton shower and hadronization effects are simulated by the \texttt{Pythia} 6.4 event generator~\cite{Sj_strand_2006}. 
The beam energy spectrum and beam-beam interaction producing incoherent \ee background pairs are generated with \texttt{Guinea-Pig}~\cite{Schulte:1998au}.
Other background sources, such as $\gamma\gamma$ to low $p_T$ hadrons, are generated separately and overlaid on the simulated events~\cite{Chen_1994}. 
A description of the whole procedure to generate all SM processes is given in \cite{Berggren:2021sju}. 
Each set of samples features fully longitudinally polarized beams in various configurations. 
Samples with realistic polarization scenarios are obtained by merging these samples with appropriate weights.

This work is based on the analysis of $e^{-}e^{+}\rightarrow\qqbar$ events at high center-of-mass energies~\cite{Irles:2023ojs,Irles:2023nee}. 
In the SM, at \cme larger than the $Z$ boson mass, these processes are sensitive to $\gamma$ and $Z$-couplings.
However, due to QED ISR, the center-of-mass energy of the \ee system may be reduced with respect to the \cme of the collider. 
Furthermore, if the energy radiated in the QED ISR  process is large enough, the \ee system may undergo a radiative return to the \Zpole, producing an on-shell $Z\rightarrow\qqbar$ process. 

The ISR changes the kinematic event properties,
so such events can be distinguished from the
high mass signal events.
It is therefore treated as a background, which we name ``$\gamma\qqbar$'' or ``radiative-return''.
For a formal separation between the signal and radiative-return processes, we define as signal those events with a quark-pair invariant mass larger than 140 GeV for ILC250 (200 GeV for ILC500) and acolinearity smaller than 0.3, with acolinearity defined as in Eq. 7 from ~\cite{Irles:2023ojs}, using a simplified definition of the acolinearity
\begin{align}
    \sin{\Psi_{acol}}=\frac{|\vec{p_{q}} \times \vec{p}_{\bar{q}}|}{|\vec{p_{q}}|\cdot|\vec{p}_{\bar{q}}|}, 
    \label{eq:acol}
\end{align}
with $\cos(\Psi_{acol}) < 0$. 

Other sources of backgrounds are due to hadronically decaying di-boson or, for ILC500, top-quark pair production.
The cross sections for all these processes, involving $q=u,d,s,c,b$ in the final state, are summarized in Tab.~\ref{tab:cross}.

\begin{table}[!ht]
\caption{Expected cross sections for signal ($\qqbar$) and main background processes. \label{tab:cross}}
\resizebox{0.48\textwidth}{!}{%
\centering
\begin{tabular}{c|cc|cc}
Process & \multicolumn{2}{c|}{ ILC250 }  & \multicolumn{2}{c}{ ILC500 } \\
$\sigma$ [pb] & $(-0.8,+0.3)$ & $(+0.8,-0.3)$ &  $(-0.8,+0.3)$ & $(+0.8,-0.3)$ \\
\hline
$\qqbar$ & 17.2 & 6.4 & 2.9 & 2.0 \\
\hline
$\gamma\qqbar$ & 60.2 & 39.3 & 16.3 & 9.5 \\
$q \overline{q} q \overline{q}$ (no Higgs) & 16.2 & 1.5 & 8.2 & 0.6 \\
$HZ \rightarrow q\bar{q}H$  & 0.2 & 0.1 & 0.07 & 0.05 \\
$b\bar{b}q_{1}\bar{q}_{2}q_{3}\bar{q}_{4}$ & -- & -- & 0.4 & 0.2 \\
\end{tabular}}
\end{table}

\subsection{Event reconstruction}
\label{sec:reco}

The ILD track reconstruction~\cite{Gaede_2014,ILD:2020qve} is based on pattern recognition algorithms carried out independently in the different parts of the tracker systems.
This is followed by the combination of all the track candidates and segments for a final refit performed with a Kalman filter. 
The entire process is implemented in the \texttt{MarlinTrk} framework, which is part of the \texttt{ILCSoft} toolkit.
The resulting tracks are combined with calorimeter hits in the Pandora particle flow algorithm (PFA)~\cite{Marshall:2015rfa} to produce a set of particle flow objects (PFO), each of which should correspond to a final-state particle.

\subsubsection{Vertex and jet reconstruction}
\label{sec:reco_vtx}

Once the PFOs are reconstructed, vertex reconstruction, jet reconstruction, and jet flavor tagging are performed using \texttt{LCFIPlus}~\cite{Suehara:2015ura}. 
The primary vertex of the event is found in a tear-down procedure, starting with all tracks and gradually removing tracks less compatible with being associated with the primary vertex hypothesis. 
In a second step, \texttt{LCFIPlus} performs an iterative reconstruction of secondary vertices. 
Jets are reconstructed using the $VLC$ algorithm~\cite{Boronat:2016tgd} for \ee colliders in exclusive two-jet mode. This algorithm includes additional beam jets which are disregarded.

\subsubsection{Flavor tagging}
\label{sec:reco_flav}

The $b$ and $c$-quark tagging are performed by \texttt{LCFIPlus} using boosted decision trees (BDTs) based on sensitive variables from tracks, vertices, and charged-hadron identification using TPC information. 
BDTs are trained using reconstructed jets in \eeqq events at $\sqrt{s}=250$ or 500 GeV accordingly~\cite{Irles:2023nee}. 
Particle Swarm Optimization~\cite{488968} with combined Kolmogorov-Smirnoff~\cite{kolmogorov1933sulla,smirnov1948table,hodges1958significance} and Anderson-Darling~\cite{cf37c5fc-d933-3771-9586-9d6b4b285d8b,doi:10.1080/01621459.1954.10501232,engmann2011comparing} tests are applied to optimize the BDTs performance while avoiding overtraining. 
The resulting BDT working points are chosen to limit the mistagging of the other quarks to $\sim1.5\%$ $(\sim 3\%)$ in the case of \bquark (\cquark) tagging.

\subsubsection{Jet-charge measurement}
\label{sec:reco_jetcharge}

The measurement of the differential cross-sections and \Afb requires measuring the jet charges, \textit{i.e.} the separation of jets originating from quarks and anti-quarks. 
Two methods are used to measure the jet-charge \cite{Irles:2023ojs}: the {\em Vertex~Charge} and {\em Kaon~Charge}.

\subsubsection*{Vertex Charge, \Bc}
\label{sec:reco_jetcharge_Bc}

The Vertex Charge is defined as the sum of the charges of all tracks
associated with reconstructed displaced vertices within a jet.
The perfect application of the method requires that all charged tracks from $b$ or $c$-hadron decays are correctly measured and associated with the jet's secondary vertex. 

The tracking efficiency in ILD for isolated tracks with momentum above 1 GeV, transverse momentum above 100 MeV and $|\costheta|<0.85$ is estimated to be close to 100\% and better than 99\% in the very forward direction~\cite{ILD:2020qve}.
However, when reconstructing secondary vertices in dense jets, loss of tracks or incorrect association with secondary vertices may not be negligible. 
It has been estimated that up to $\sim5\%$ of the tracks may be missed during vertex reconstruction, especially in the very forward and backward regions~\cite{Irles:2023ojs}.
This is particularly relevant for \bquark jets due to the large number of secondary tracks per jet: most jets have between two and five (and in rare cases up to fifteen) displaced tracks.
In contrast, most \cquark jets contain two displaced tracks.
The difference in performance for the two flavors can be qualified by $P_{chg}$, the probability that the jet charge reproduces the sign of the charge of the initial quark. 

\subsubsection*{Kaon Charge, \Kc}
\label{sec:reco_jetcharge_Kc}

The charge of the jet can also be estimated by 
the charge sum of all kaons associated to displaced vertices inside the jet.
These kaons are produced in decays of $D$-mesons and $B$-hadrons. 
The performance of this method is limited mostly by charged-kaon identification capabilities and physics processes such as $B^0 - \bar{B}^0$ oscillations.
The baseline ILD design foresees a $K^{\pm}/\pi^{\pm}$ separation power of $\sim$3\footnote{The separation power is calculated as defined in Eq. 8.3  of Ref.~\cite{ILD:2020qve}} but an improvement is expected when using alternative TPC designs with larger pixelation allowing cluster-counting reconstruction, \dNdx, in place of the traditional \dEdx~\cite{LCTPC:2022pvp}.
We implemented this feature of improved reconstruction in our analysis, such that the TPC track reconstruction offers $K^{\pm}/\pi^{\pm}$ separation power up to 4 using \dNdx measurements for tracks with momentum between $\sim(3-30)$ GeV. A detailed description of the discriminating variables and procedure can be found in \cite{Irles:2023nee} where it is shown that using \dEdx for single charged-kaon identification in displaced tracks inside jets, we can get up to an $\sim80\%$ efficiency with $\sim80\%$ purity for the \bbbar case at ILC250 and ILC500, an $\sim80\%$ efficiency with $\sim90\%$ purity for the \ccbar case at ILC250 and an $\sim80\%$ efficiency with $\sim80\%$ purity for the \ccbar case at ILC500 (because of the higher momentum of displaced tracks compared with ILC250). However, for the \dNdx at the ILC250 scenario, we can work at the points with $90\%$ efficiency and $\leq95\%$ for both final states. At ILC500, the purity is slightly worse due to the higher momentum of the displaced tracks but still above the $90\%$. 


\subsubsection{Event preselection}
\label{sec:reco_preselection}

Once events are reconstructed, a preselection is based on a series of kinematic cuts applied to enrich the data sample of \qqbar ($q=u,d,s,c,b$) signal events, expected to show back-to-back two-jet kinematics, while removing the backgrounds. 
The largest background contamination is the radiative-return background associated with QED ISR.
Most QED ISR will be collinear to the beam and, if energetic enough, will result in a reconstructed invariant mass of the two-jet system significantly smaller than $\sqrt{s}$, making it straightforward to filter such events.
However, a non-negligible fraction of the radiative-return events, $\sim10\%$ according to the existing simulations, will radiate photons inside the detector volume, requiring other kinematic variables, defined in the following, to be used to filter these events.

We define two new objects, $\gamma_{clus,j}$, by clustering all uncharged PFOs within each jet ($j$). 
We denote the most energetic of these two as $\gamma_{clus.}$.
The reconstructed energy and angles of $\gamma_{clus.}$ will be used for the event selection.
Another variable used is $y_{23}$, the distance (as defined by the $VLC$ algorithm) at which the event transitions from a two to a three-jet topology.
We also use the invariant mass of the two-jet system ($m_{jj}$).
The selection requirements for ILC250 are summarized in the following, with the modifications for ILC500 shown inside parentheses: 
\begin{enumerate}
    \item photon veto cuts, rejecting events if:
    \begin{enumerate}
        \item at least one of the jets contains only one PFO;
        \item if the energy of $\gamma_{clus.}$ is larger than 115 GeV (220 GeV) or it is reconstructed in the forward region $|\costheta|>0.97$;
    \end{enumerate}
    \item events with $(\sin \Psi_{acol})_{jj}$ > 0.3 are rejected;
    \item events with $m_{jj}$ < 140 GeV (200 GeV) are rejected;
    \item events with $y_{23}$ > 0.02 (0.007) are rejected.
\end{enumerate}
Cuts 1 to 3 are designed to reduce the contamination from the radiative-return backgrounds, and cut 4 is most effective against di-boson backgrounds. 
The last cut is significantly tightened at ILC500 to reduce the much larger $WW$ background contamination in the case of left-handed electron beam polarization. 
This cut reduces the efficiency from $\sim75\%$ to $\sim55\%$ while keeping the background over signal $B/S$ ratio below $3-5\%$. 
At ILC500, cut 2 (on acolinearity) reduces the \ttbar contribution to the 1\% B/S level, and cut 4 further reduces the \ttbar\ background to negligible levels.

\todo{The event selection has been thoroughly studied and documented in Refs.~\cite{Irles:2023ojs,Irles:2023nee}. According to these studies and the references cited, a constant within the percent-level or less high-efficiency rate is obtained in the entire detector volume, except for the very forward/backward region of $|\cos\theta|>0.9$, where the efficiency rapidly drops. This drop in efficiency is due to the removal of events with high-energy photon candidates in that region in order to eliminate radiative-return events, as well as the decrease in tracking and vertexing performance, which impacts the flavor tagging algorithms. In Ref. \cite{Irles:2023ojs}, it was cross-checked that, within current uncertainties and using leading-order simulations, the difference between extrapolating over or ignoring that region is negligible and does not affect the analysis. In the present analysis, we have followed the former alternative. However, we emphasize the importance of the optimization of detector layout and reconstruction tools, especially in the forward/backward region. This is currently being discussed within the ILC Concept Group as well as within the Higgs Factory detector concepts~\cite{ILD:2020qve,Bilokin:2017ody,Mizuno:2020,deBlas:2024bmz}.}

\section{Experimental reconstruction of the forward-backward observable using full simulation tools at ILC250 and ILC500}
\label{sec:AFB}

This section describes the most critical aspects of the experimental reconstruction of the observable forward-backward asymmetry \Afbq by measuring $\frac{d\sigma}{d\costheta}$.
This work is based on a previous ILD study~\cite{Irles:2023ojs,Irles:2023nee}. 
The method starts with a preselection that results in a highly pure \qqbar sample, followed by the double flavor tagging (Sec~\ref{sec:AFB_DT}) which selects \bquark (or \cquark) events, and ends with the double charge (Sec.~\ref{sec:AFB_DC}) measurement to distinguish between quark and anti-quark jets.
The differential cross-section is extracted from the measurement of the total number of events reconstructed as a function of the quark-jet scattering angle $\theta$:
\begin{eqnarray}
    \frac{dN}{d\costheta} &=& \mathcal{L}\left[\varepsilon_{pre} \varepsilon_{DTC} \frac{d\sigma}{d\costheta} +
    \varepsilon_{bkg} \frac{d\sigma_{bkg}}{d\costheta}\right]
    \label{eq:N_AFB}
\end{eqnarray}
where $\mathcal{L}$ is the integrated luminosity and the $\varepsilon$ variables are the different selection efficiencies, defined as a function of $|\costheta|$.
The signal preselection efficiency ($\varepsilon_{pre}$) and the background selection efficiency ($\varepsilon_{bkg}$) are estimated using only Monte Carlo data.

\subsection{Double Tagging method}
\label{sec:AFB_DT}

The Double Tag method (\DT) \cite{ALEPH:2005ab} is based on the comparison of single and double flavor-tagged samples for the simultaneous extraction of the tagging efficiency, $\epsilon_{q}$ and the hadronic cross-section fraction \Rq\footnote{The hadronic cross-section fraction is defined as \begin{equation}
  \Rq= \frac{\sigma_{\eeqq}}{\sigma_{had.}}
\end{equation}
where $\sigma_{had.}$ is defined as $\sigma_{\eeqq}$ integrated over all quark flavors except the top-quark.}.
The method is applied once we have preselected an enriched \qqbar sample reconstructed as two jets. 
We next apply flavor tagging to all jets, and extract two numbers:
the fraction of all jets tagged as being of flavor $q$, and the fraction of events in which both jets are tagged as flavor $q$.
The exact formulation is described in Ref.~\cite{Irles:2023ojs} for a fully differential analysis, in contrast to the integral analyses performed in the past. 
By comparing these two ratios for \bquark and \cquark, we can simultaneously measure the 
efficiency of the flavor tagging algorithm ($\epsilon_{q}$) and \Rq for both flavors. 

Although the method is based on data comparisons, some initial hypotheses, based on simulations, are required.
For instance, it is assumed to be an almost background-free analysis (or to have a perfectly modeled background). This was easier to achieve at LEP and SLC running at the \Zpole. 
For data taken in the continuum above the \Zpole, a tighter preselection is required to minimize the background contribution of the radiative return or di-boson hadronic decays. 
The method also assumes knowledge of the mis-tagging efficiencies, the probability of tagging a true $q^\prime$ as $q$. 
The size of these mis-tagging efficiencies and their uncertainties will directly impact the uncertainty of the measurements. 
This factor was one of the dominant sources of uncertainty for the measurements at LEP and SLC, with less relevance in the latter case due to its improved flavor-tagging capabilities (especially for the \cquark). 
Finally, the method assumes that the quark-tagging efficiencies are symmetric between the two sides of the detector (positive and negative \costheta). 
However, effects such as the resolution on the primary vertex reconstruction, inhomogeneities in the detector layout or performance, or kinematic variations of the back-to-back topology due to hard gluon radiation may introduce correlations. 
These correlations can be parametrized by (1+$\rho_q$), known in the literature as the hemisphere or jet angular correlation. With this notation, the double-tag efficiency is given by $\epsilon_{q}^{2}\cdot(1+\rho_q)$. This parameter is the only one that is calculable only using Monte Carlo simulations, and it was considered a significant source of uncertainty in past experiments. 
It was less of an issue at SLC than at LEP thanks to the more precise primary vertex determination.
The jet angular correlation is expected to be negligible in the case of the ILD at ILC~\cite{Irles:2023ojs} except in the forward detector region, where the tagging efficiency drops due to limited acceptance.
In a fully differential analysis, these regions and others affected by detector issues can be removed from the study when performing a fit and be substituted by extrapolations. 

The preselection and \DT selection efficiencies for the two flavors are shown in Tab.~\ref{tab:yield} for the ILC250 and ILC500. 

\subsection{Double Charge method}
\label{sec:AFB_DC}

The Double Charge method (\DC) requires a pure \qqbar sample (with $q$ being only one flavor, $c$ or $b$) for its application. 
It consists of the measurement of the charge of each jet using one of the methods described in Sect.~\ref{sec:reco_jetcharge}. 
Only events with two oppositely charged jets are accepted.
The experimental determination of the probability \Pb that the jet charge reproduces the sign of the charge of the quark of the hard scattering is straightforward~\cite{Irles:2023ojs}. 
Since the \Bc and the \Kc show similar values of \Pb, it is also possible to use mixed cases in which opposite jets use different methods.
In Ref.~\cite{Irles:2023ojs}, three different categories are defined and studied for each flavor: 
\todo{
\begin{enumerate}
    \item The charge of both jets has been measured with the method that more often gives a non-zero charge measurement (primary method).
    \item Only one jet has no measurement of its charge with the primary method but does have it with the other method (secondary method).
    \item Both jets have their charge measured with the secondary method.
\end{enumerate}}

\todo{For the \bquark, the primary method is the \Bc, while for the \cquark case it is the \Kc .
The algorithm defined in Ref.~\cite{Irles:2023ojs} has the benefit that it does not rely on the determination of the efficiency of each method since it only uses the measured fractions, $f$, of jets with non-zero charge measurements and the probability of having correctly measured the sign of the charge. 
For the latter, we use the \Pb, first introduced in Sect.~\ref{sec:reco_jetcharge}, measured using the data sample itself by comparing events with the same sign of the charge measurements for both jets and events with different signs. With simple combinatorics calculations, the experimental \Pb values will be extracted without the need for Monte Carlo simulations. This procedure is performed differentially, in small bins of \costheta.}

\begin{table}[!ht]
\caption{\todo{Jet Charge performance of the different methods showing the fraction, $f$, of jets with non-zero charge measured by each method and the probability of having correct measurement, \Pb, as described in the text and in Ref.~\cite{Irles:2023ojs}. For simplicity, the values are given only for events excluding the forward regions $|\costheta<0.9|$. \label{tab:DC}}}
\centering
\begin{tabular}{c|cc|cc}
   & \multicolumn{2}{c|}{ILC250} & \multicolumn{2}{c}{ILC500}\\
\hline
 & $c$ & $b$ & $c$ & $b$ \\
 \hline
$f_{Vtx}$   & 36$\%$ & 65$\%$ & 49$\%$ & 75$\%$ \\
$f_{K}$     & 49$\%$ & 48$\%$ & 51$\%$ & 72$\%$ \\
$P_{chg,Vtx}$ & 93$\%$ & 83$\%$ & 92$\%$ & 82$\%$ \\
$P_{chg,K}$   & 95$\%$ & 75$\%$ & 80$\%$ & 70$\%$ \\
\end{tabular}
\end{table}

\todo{The \DC measurements minimize the migrations between hemispheres due to wrongly measured jet charge. 
However, these migrations are not negligible in the \bquark with high left-handed electron-beam polarization case, even using the \DC. 
The reason is two-fold: the relatively small \Pb for the \bquark case (as explained in Sec. \ref{sec:reco_jetcharge}) and the large value of \AFB that produces a larger amount of event migration from the forward to the backward region than from backward to forward. 
This has a sizable impact on the reconstructed differential distribution, with localized deviations of the differential cross-section of up to a factor of three in the region $\costheta\lesssim -0.75$.
These migrations can be easily corrected using the measured values of \Pb, which give the probability that the charge has been wrongly measured in both jets, hence flipping the sign of \costheta~\cite{Irles:2023ojs}.}

\todo{The efficiency of the full selection, including preselection, \DT, and \DC selection for the two flavors, are shown in Tab.~\ref{tab:yield} for the ILC250 and ILC500 cases, for the full detector volume, \textit{i.e.,} including the forward regions.}

\begin{table}[!ht]
\caption{Efficiencies and $B/S$ estimated using ILD full simulation and reconstruction.\label{tab:yield}}
\resizebox{0.48\textwidth}{!}{%
\centering
\begin{tabular}{c|cc|cc}
 & \multicolumn{4}{c}{Eff(B/S)$[\%]$} \\
  & \multicolumn{2}{c|}{ILC250$(-0.8,+0.3)$} & \multicolumn{2}{c}{ILC250$(+0.8,-0.3)$}\\
\hline
Preselection:  & \multicolumn{2}{c|}{ $udscb:$ 75.8(5.6) }& \multicolumn{2}{c}{ $udscb:$ 75.9(4.8) } \\
+\DT  & $c:$ 9.5(0.6) & $b:$ 38.1(0.9) & $c:$ 9.4(0.5) & $b:$ 38.1(1.3) \\
+\DC  & $c:$ 3.7(0.5) & $b:$ 14.7(0.8) & $c:$ 3.7(0.5) & $b:$ 14.7(1.2) \\
\hline
\hline
& \multicolumn{2}{c|}{ILC500$(-0.8,+0.3)$} & \multicolumn{2}{c}{ILC500$(+0.8,-0.3)$}\\
\hline
Preselection:  & \multicolumn{2}{c|}{ $udscb:$ 54.4(7.4) }& \multicolumn{2}{c}{ $udscb:$ 54.5(5.0) } \\
+\DT  & $c:$ 10.9(0.3) & $b:$ 29.7(0.3) & $c:$ 10.8(0.3) & $b:$ 29.5(0.3) \\
+\DC  & $c:$ 3.2(0.3) & $b:$ 12.1(0.3) & $c:$ 3.1(0.5) & $b:$ 12.1(1.3) \\
\end{tabular}}

\end{table}

\begin{figure*}[!ht]
    \centering
    \begin{tabular}{cc}
            \includegraphics[width=0.45\textwidth]{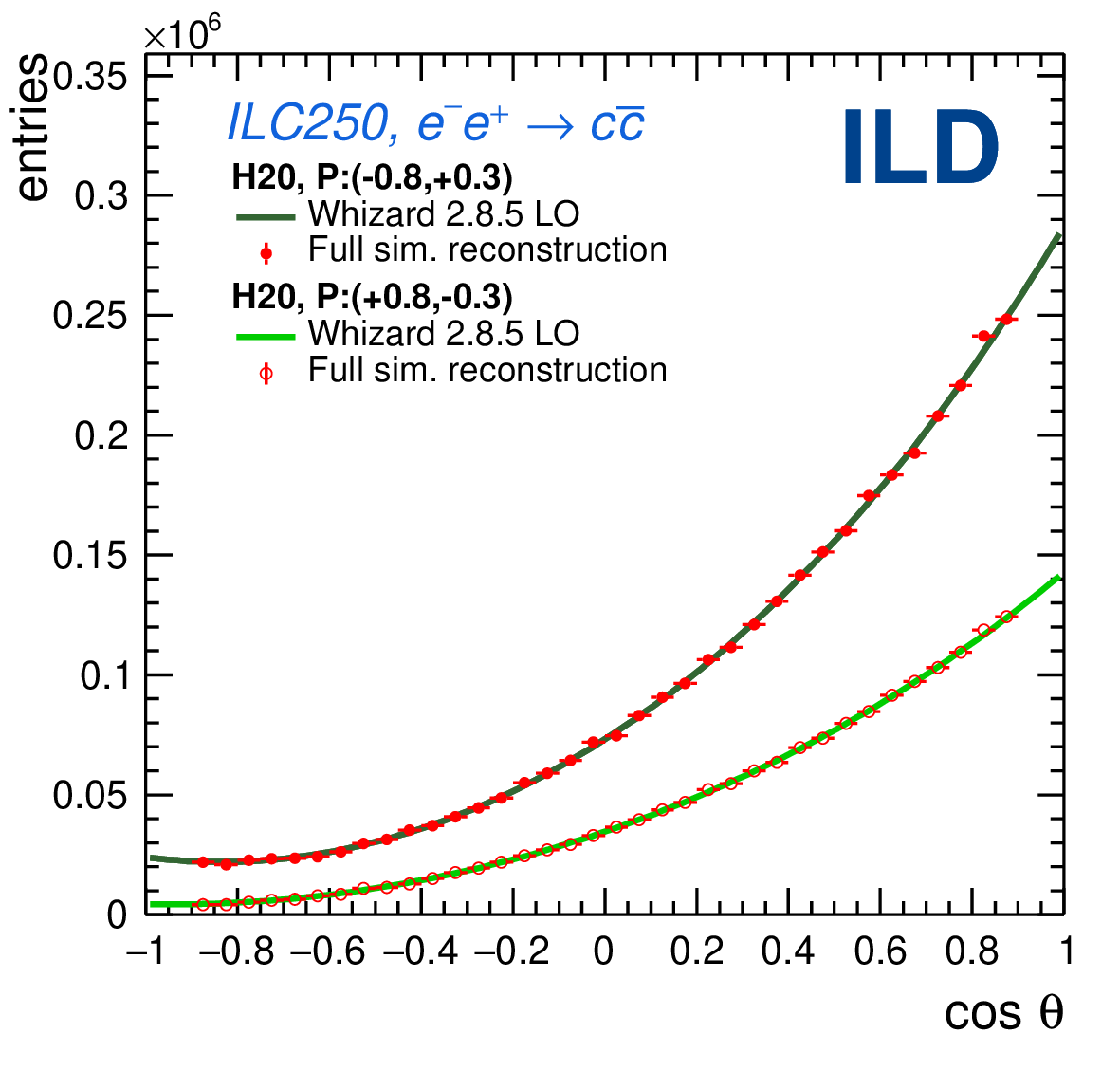} & 
            \includegraphics[width=0.45\textwidth]{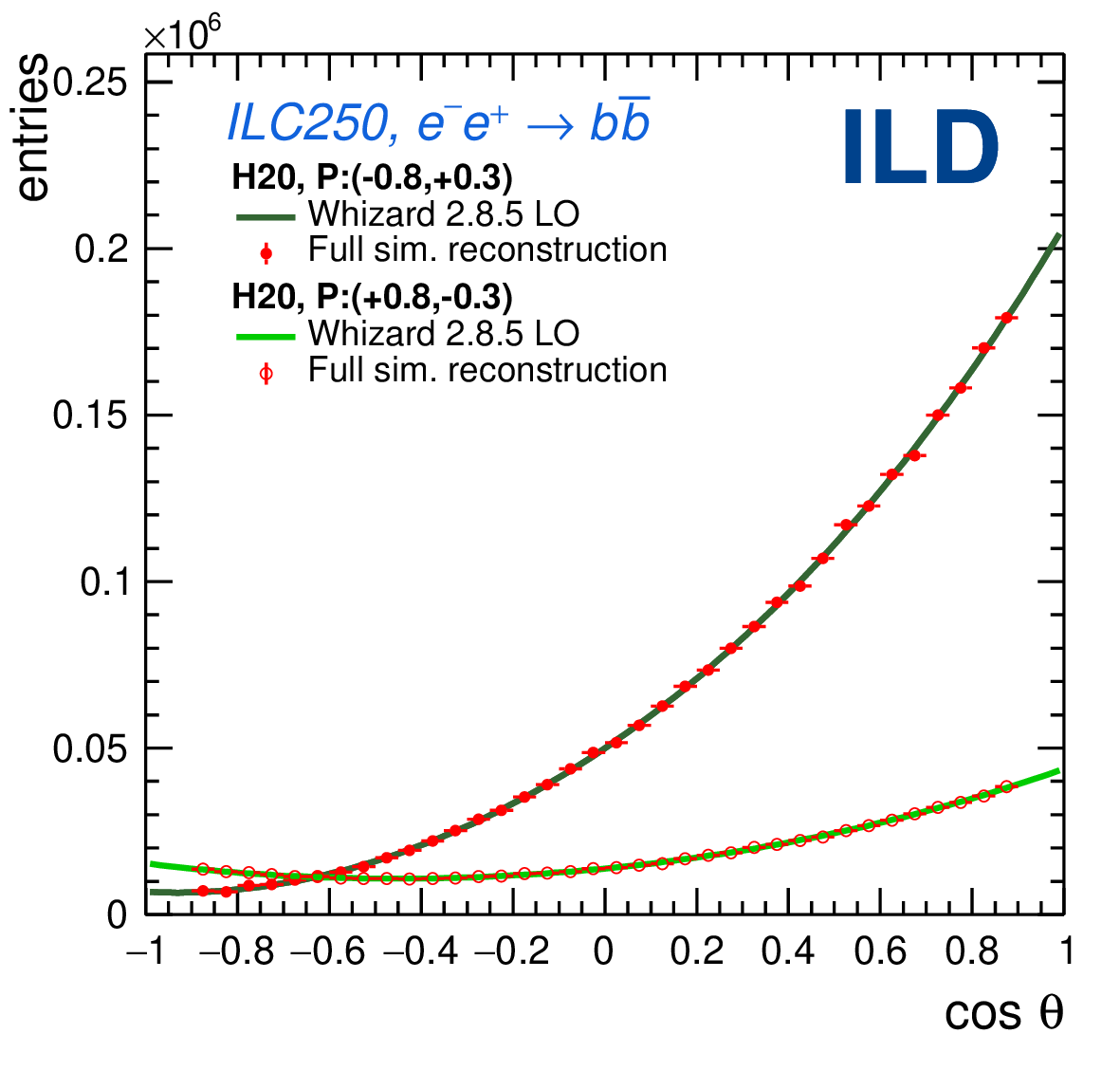}  \\
            \includegraphics[width=0.45\textwidth]{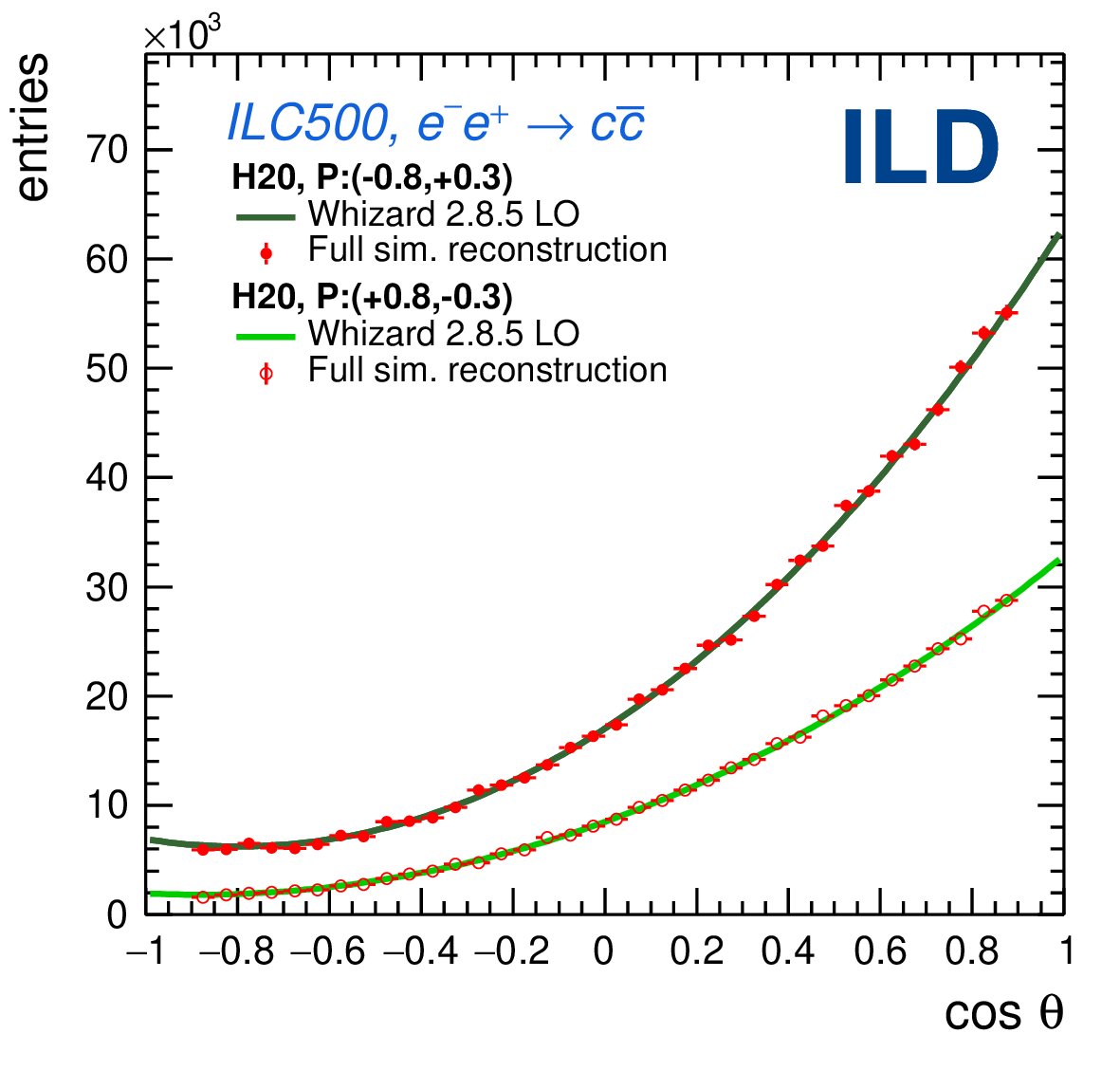} & 
            \includegraphics[width=0.45\textwidth]{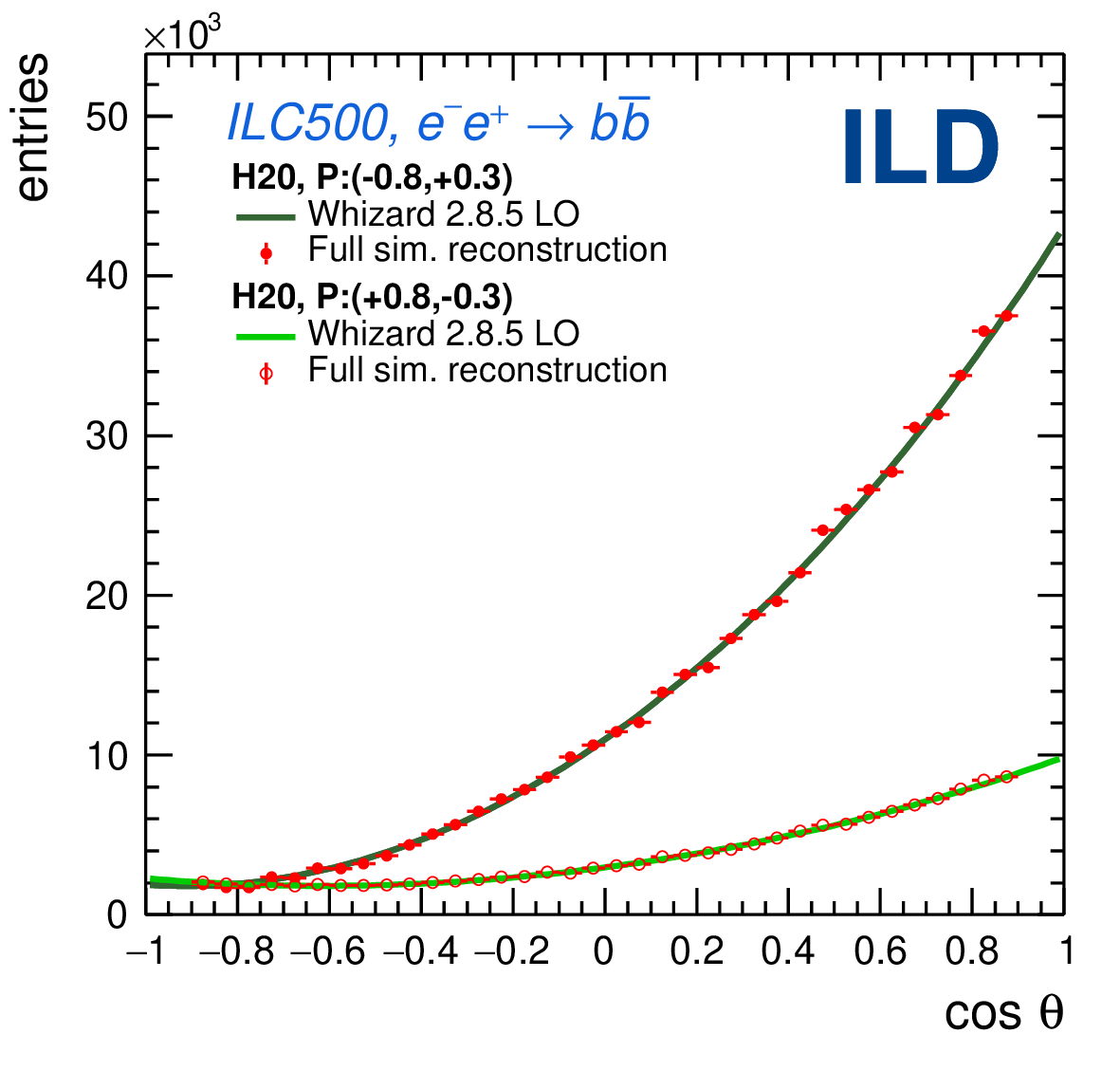}  
    \end{tabular}
    \caption{Comparison of parton level and fully reconstructed 
    $d\sigma/d\costheta$
    distributions for different ILC scenarios and the $b$ and \cquark. }
    \label{fig:AFBfit}
\end{figure*} 

\subsection{Estimation of \AFB}
\label{sec:AFB_AFB}

Once the \DT and \DC methods have been applied, all the inputs needed to extract the efficiency of double tagging and charge measurement, $\varepsilon_{\DTC}$, and to infer $d\sigma/d\costheta$ are available. 
The efficiency is estimated for every flavor and each \DC category, and the statistical uncertainties are propagated accordingly~\cite{Irles:2023ojs}. 
The correction for the efficiencies is performed differentially as a function of $|\costheta|$.

The result of the entire correction procedure is shown in Fig.~\ref{fig:AFBfit} for the different flavors and running scenarios using SM samples.

\begin{figure}[!ht]
    \centering
            \includegraphics[width=0.45\textwidth]{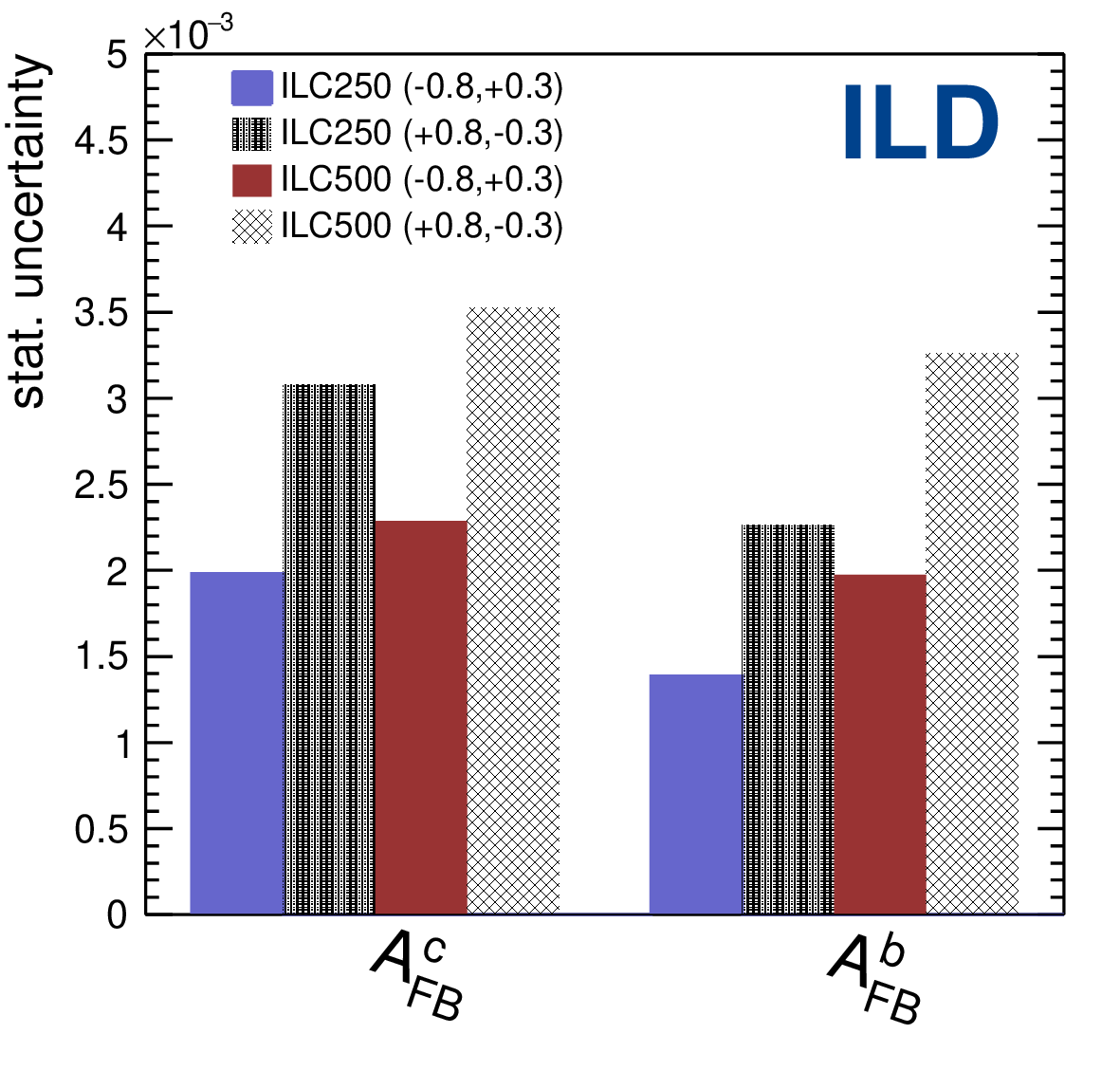} 
    \caption{Estimated statistical uncertainties on \AFBc and \AFBb using ILD full simulation and reconstruction at ILC250 and ILC500.}
    \label{fig:AFBstat}
\end{figure}

For the estimation of \Afb, the function
\begin{flalign}
    \frac{d\sigma}{d\costheta} = S \left(1+\text{cos}^{2}\theta\right) + A  \costheta
    \label{eq:fiteq}
\end{flalign}
is fit to the reconstructed distributions. 
Note that this function neglects the SM tensorial contribution $T\times \text{sin}^{2}\theta$, which is very small given the large boost of $b$ and $c$-quarks. 
The fit is performed in the range of $|\cos\theta| < 0.9$, avoiding the very forward regions in which the acceptance decreases.
The \AFB is extracted by extrapolating the fitted function to the full range of \costheta, and the statistical uncertainty is estimated for the expected integrated luminosity, including the uncertainties of the correction methods.
The expected statistical uncertainties for the various running scenarios are shown in Fig.~\ref{fig:AFBstat}.

\subsection{Systematic uncertainties}
\label{sec:AFB_syst}

At ILC250 and ILC500, the expected statistical uncertainties on \AFB are at the level of a few per mil. 
A comprehensive study of the leading experimental systematic uncertainties
is reported in Ref.~\cite{Irles:2023ojs}.
The most significant systematic uncertainties on \AFB are due to the preselection efficiency, the hadronization/fragmentation modeling, the angular correlations due to QCD effects, and the knowledge of the beam polarization.
These are reported in the following. However, we emphasize that the expected size of these systematic uncertainties is negligible
compared to the statistical uncertainties expected at ILC when running above the \Zpole.

\subsubsection{Pre-selection efficiency}

Although this efficiency cancels in the numerator and denominator of the integral calculation of \AFB, 
the preselection efficiency cannot be neglected in a full differential analysis since it affects the shape of the differential measurement. 
The impact of this uncertainty has been evaluated by producing pseudo-data distributions applying uncorrelated 10$\%$ relative variations in bin-by-bin preselection efficiencies.
This uncertainty is propagated to \AFB, giving rise to a relative uncertainty on \AFB of $\leq 0.1\%$.
Uncertainties from background mis-modeling become mostly negligible, $\simeq 0.01\%$, due to the very efficient background rejection.

\subsubsection{Hadronization/fragmentation modeling}

\todo{Uncertainties on \AFB related to fragmentation are expected to be negligible 
thanks to the \DT and \DC methods. These were proposed and used by LEP and SLC experiments. These methods rely on the data for the estimation of the flavor tagging efficiencies, minimizing the usage of Monte Carlo tools for the modeling. Hence, fragmentation model uncertainties will only affect to the mis-tagging efficiency estimation but not the tagging efficiency.}
Furthermore, the mis-tagging rates will be much lower than at past experiments thanks to the high performance expected for modern flavor-tagging algorithms using modern statistical and machine-learning techniques and the progress in detector technologies. 

\subsubsection{Angular correlations}

Full simulation studies suggest that the value of $\rho_q$ at ILC250 is smaller than 0.2\% throughout most of the detector volume~\cite{Irles:2023ojs}.
The tracking system in the ILD simulation is symmetric, and no coherent noise is simulated, indicating that a non-zero $\rho_{q}$ value can only be the result of occasional mis-measurements of the primary vertex or hard QCD radiation diluting the back-to-back configuration of the di-jet system. 
The small value of $\rho_{q}$ suggests that both effects can be effectively controlled. 
This results from the small beam size (jet angular correlations due to a misplaced common vertex are suppressed) and an excellent tracking system. 
Moreover, a high tagging efficiency can significantly reduce jet angular correlations. 
To consider the impact of angular correlations due to QCD radiation, we assume they contribute an uncertainty of $\lesssim0.1\%\cdot\AFB$, following Ref.~\cite{AlcarazMaestre:2020fmp}, after having introduced acolinearity cuts in our definition of the signal and selection procedure. 
A full assessment of this effect would require simulations based on NLO QCD.

\subsubsection{Beam polarization}

Beam polarization uncertainties \cite{Karl:424633} influence the accuracy of precision measurements.
For the measurement of \AFB, this uncertainty affects the \bquark and \cquark flavors differently, and has a non-negligible impact only on the right polarization scenario for the b-quark;
in this scenario, we expect an uncertainty contribution of $0.15\% \cdot \AFB$ at ILC250, and somewhat smaller at ILC500~\cite{Irles:2023ojs}.

\section{Statistical discrimination power for GHU models at ILC}
\label{sec:analysis}

The detailed studies described in the previous section result in a realistic estimation of the uncertainties on \AFB for the $b$ and \cquark at ILC250 and ILC500, with existing detector models and reconstruction tools.
We have shown that the statistical uncertainties on \AFB expected at ILC running above the \Zpole
are much larger than any systematic effects, which are therefore ignored in this section.
Furthermore, we assume Gaussian uncertainties and uncorrelated measurements. 
This second approximation is motivated by the nature of the analysis, in which the \DT and \DC methods lead to the selection of fully independent samples for the different flavors and beam polarizations. 
Statistically independent MC simulations have been used to analyze the various polarization scenarios.

To probe the potential of ILC to indirectly search for new physics, the significance when comparing two models, $i$ and $j$, is defined as
\begin{equation}
    d_{ij}=\frac{|A_{FB,i}-A_{FB,j}|}{\Delta A_{FB,j}}
\end{equation}
with $A_{FB,i/j}$ being the \AFB predicted at leading order by model $i$ or $j$, as explained in Sect.~\ref{sec:theory}. 
The $\Delta A_{FB,j}$ corresponds to the expected 
statistical uncertainty of the forward-backward measurement at ILC, obtained as explained in Sect.~\ref{sec:AFB}.
In addition, the systematic uncertainties arising from the limited knowledge of the exact value of the couplings in the SM prediction are added in quadrature to $\Delta A_{FB,j}$. 
For these estimates, we used different assumptions on the uncertainties on the Z-boson couplings to fermions and propagated them to the leading order calculation of \AFB. 
The different assumptions on these uncertainties extracted from Refs.~\cite{ParticleDataGroup:2020ssz,ILCInternationalDevelopmentTeam:2022izu} are:\newline
\textbf{C}: current uncertainties, from LEP and SLC measurements; \newline
\textbf{R}: expected uncertainties from measurements of radiative return events at ILC250; \newline
\textbf{Z}: expected uncertainties from measurements at a dedicated ILCGigaZ run. \newline

The probability for each $d \geq d_{ij}$ case and the discovery power discrimination are calculated in terms of the number of standard deviations from the null hypothesis for the model $i$. 
This is performed for each measurement and combined following the multivariate Gaussian formalism.

\begin{figure}[!ht]
    \centering
            \includegraphics[width=0.45\textwidth]{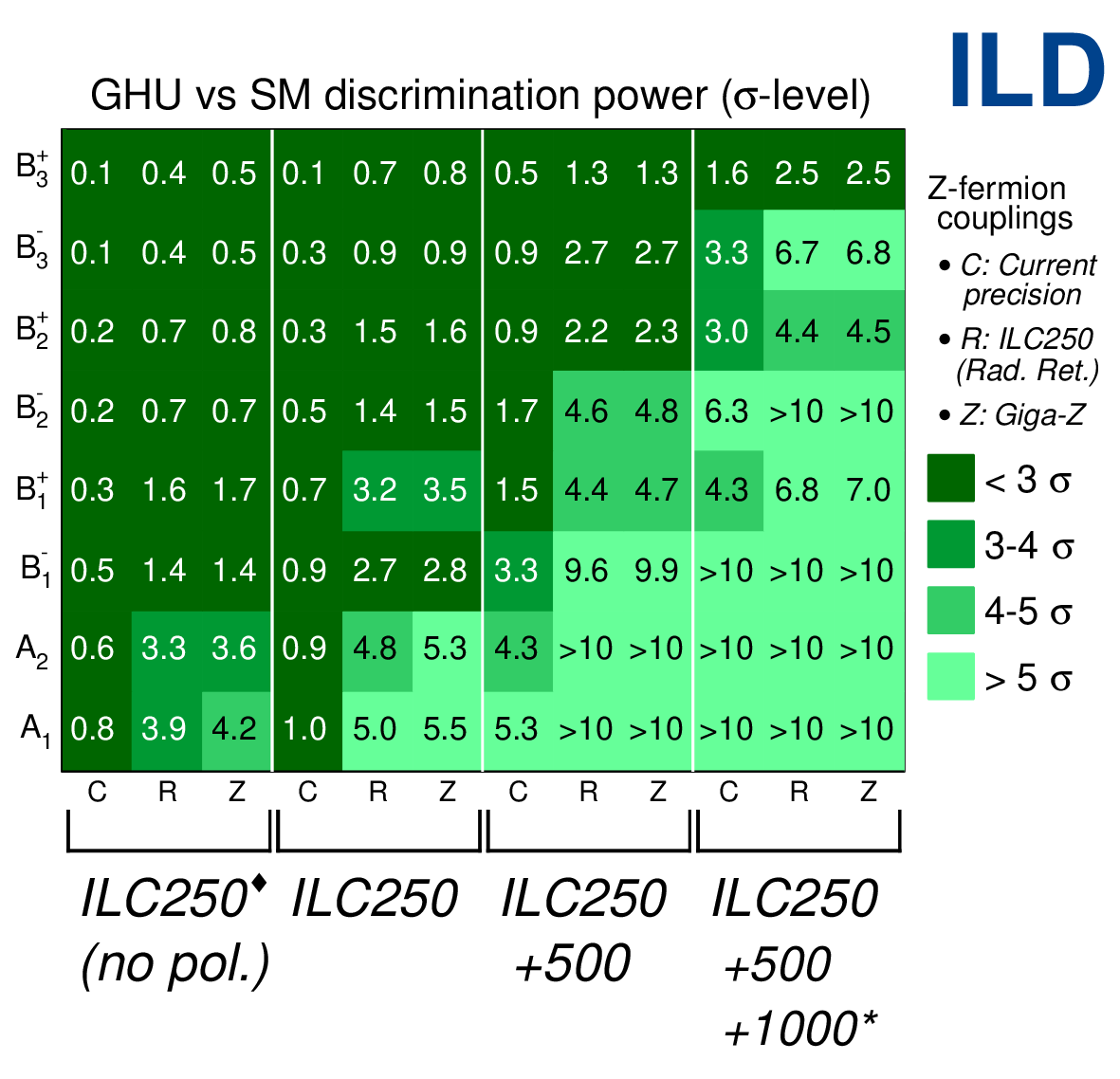}
    \caption{Statistical discrimination power between the GHU models described in the text and in \cite{Funatsu:2017nfm,Funatsu:2020haj,Funatsu:2023jng} and the SM. Different running scenarios of ILC are compared: \ILCnopol (hypothetical case with no beam polarization and 2000 \fb of integrated luminosity), ILC250 (2000 \fb), ILC500 (4000 \fb), and ILC1000* (8000 \fb, not using full simulation studies but extrapolations of uncertainties from ILC500). Three different assumptions for the $Z$-fermion couplings uncertainties are considered \cite{ILCInternationalDevelopmentTeam:2022izu}: \textit{C} for current knowledge; \textit{R} for expected knowledge after the full ILC250 program and the study of $Z$-fermion couplings from radiative return events, and \textit{Z} for expected knowledge after a full ILCGigaZ program. }
    \label{fig:SM_vs_GHU}
\end{figure}

Fig.~\ref{fig:SM_vs_GHU} shows the expected discrimination power
between the different GHU models and the SM for the different ILC running stages above the \Zpole foreseen by the H20-staged plan, ILC250 and ILC500. We have included two extra cases: 
\begin{itemize}
    \item \ILCnopol: a hypothetical case of an ILC250 operating with un-polarized beams and assuming a total integrated luminosity equal to the baseline ILC250 scenario (2000 \fb). 
    The ($^\blacklozenge$) symbol is used to distinguish this case from the other studies, which use the nominal ILC beam polarization. 
    This scenario can be compared with current circular collider proposals. Still, one should bear in mind that the simulated beam conditions are those of ILC and that the simulated ILD model has been optimized for the full range of ILC energies. Dedicated studies with full simulation of the beam conditions and detector models designed for circular colliders would be required to extract more definitive conclusions.
    \item ILC1000*: we have extrapolated statistical uncertainties from ILC500 to 8000 \fb at ILC1000, split according to the different polarization conditions as described in Tab. \ref{tab:lumNEW}. The (*) symbol is used to distinguish this case from the three other cases, which use full simulation samples and reconstruction. At ILC1000, compared with ILC250 and ILC500, the experimental challenges are expected to be slightly different, with much more collimated jets and vertices and the possible presence of new backgrounds. However, no significant differences are expected in overall detector performance since the ILD has already been optimized to operate at ILC1000 (see \cite{Behnke:2013lya,ILCInternationalDevelopmentTeam:2022izu} and references therein).
\end{itemize}

\subsection{Requirements on the precision of the $Z$-fermion couplings measurements}
\label{sec:analysis_other_gigaz}

Even if large deviations in the \AFB observables are measured at ILC250 or at higher energy in \eeqqbar, one could still be unable to distinguish the contribution of new resonances from deviations from the SM $Z$-boson couplings. 
For this reason, high-precision measurements of the fermionic $Z$ couplings are required.
An ILCGigaZ run would allow the precise determination of all $Z$-boson couplings to fermions (except the top quark). 
Full simulation experimental studies are yet to be performed at the ILC and other Higgs Factories. 
First studies~\cite{Irles:2019xny,ILCInternationalDevelopmentTeam:2022izu} show that an improvement of up to two orders of magnitude (including systematic uncertainties) could be obtained at the ILC for these couplings compared with the LEP and SLC. 
These studies include discussions of the most significant systematic uncertainties, which are expected to be 2-10 times larger than the statistical ones. 
Moreover, studies presented in Ref.~\cite{ILCInternationalDevelopmentTeam:2022izu} suggest that even at ILC250, such couplings could be measured with about one order of magnitude higher precision than at LEP/SLC by studying the radiative return to the \Zpole.

The impact of assuming improved precision on these couplings, according to the scenarios explained above, is shown in Fig.~\ref{fig:SM_vs_GHU} in the different columns $C$, $R$, and $Z$. 
It shows that at ILC250, the precision
on the $Z$ couplings from a study of radiative return events allows one
to approach the five standard deviations ($\sigma$) level for the discrimination of \Amodels vs SM and get almost  $3\sigma$ for the $B_1^\pm$. 
With ILCGigaZ precision, the discrimination power would be enhanced, allowing full discrimination of the \Amodels.
When including the higher energy stages of ILC, the difference between the two possible scenarios for improved $Z$ coupling precisions becomes less critical due to the larger size of the new resonance contributions.



\subsection{Importance of beam polarization and high energy reach}
\label{sec:analysis_pol_energy}

The ILC offers high-energy beams with a high degree of longitudinal polarization.
This provides direct access to the different helicity amplitudes at different energies. 
This is particularly important for models predicting deviations in the right-handed electroweak couplings, which are less constrained by existing measurements. 
The comparison between the first columns of Fig.~\ref{fig:SM_vs_GHU}, shows that the high degree of electron and positron beam-polarization at ILC250 offers a gain comparable to a factor two in integrated luminosity with respect to \ILCnopol.
Fig.~\ref{fig:theory1} shows that most of the sensitivity at ILC250 comes from the configuration with right-handed electron beam and left-handed positron polarization.
An increase of statistics of this run from 900 \fb to $\sim2.5\times900$ \fb would provide $5\sigma$ discrimination power for the $\Aone$, $\Atwo$ and the $B_1^\pm$ GHU models (predicting $m_{KK}=8.81,~10.3$ and $13$ TeV respectively).

\begin{figure}[!ht]
    \centering
            \includegraphics[width=0.45\textwidth]{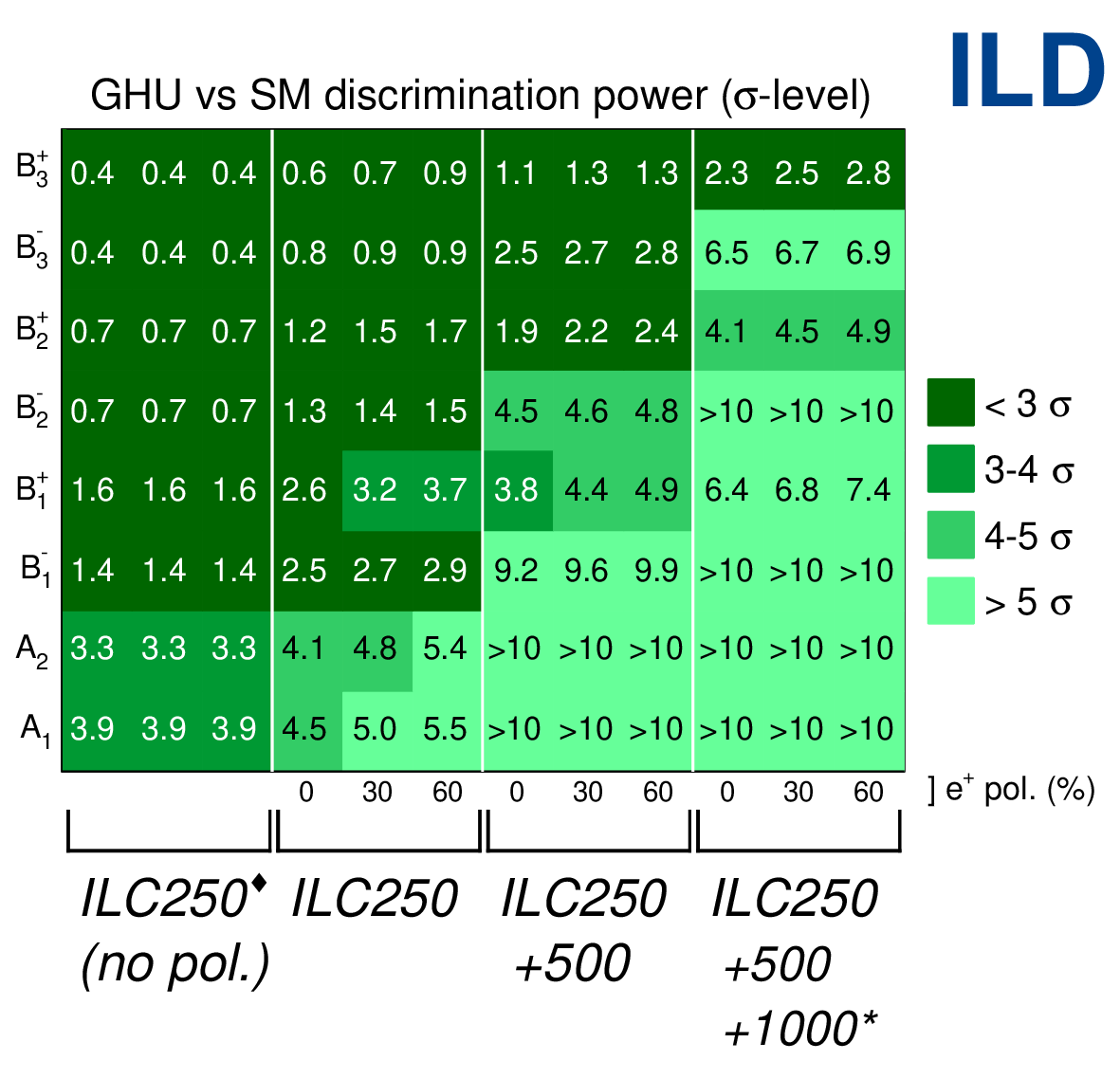}
    \caption{Similar to Fig.~\ref{fig:SM_vs_GHU} but fixing the \textit{R} case for the $Z$-fermion couplings. Three scenarios for the positron beam polarization are shown. 
    For the \ILCnopol we continue assuming no polarization for any beam.
    \label{fig:SM_vs_GHU_pol}}
\end{figure}

The production of an intense electron beam with a high degree of polarization is expected to be technically possible, while the positron source poses some technological challenges. To face these challenges, two options are being considered \cite{PositronWG,ILCInternationalDevelopmentTeam:2022izu}. 
The baseline undulator-based positron source allows the production
of polarized positron beams, while the other electron-driven concept would provide un-polarized positron beams. 
The baseline option provides $30\%$ positron beam polarization at ILC250 and ILC500 and $20\%$ at ILC1000.
An upgrade could increase the polarization to 60$\%$.
We perform the exercise of comparing several scenarios for the positron beam polarization. The result is shown
in Fig.~\ref{fig:SM_vs_GHU_pol} where we compare the discrimination power assuming three positron beam polarization scenarios: 0$\%$, 30$\%$ and 60$\%$\footnote{For simplicity we use $30\%$ positron beam polarization also for the middle column for the ILC 1 TeV case. However, it is not the baseline configuration, which foresees $20\%$ positron beam polarization.}, all of them using extrapolated uncertainties from the \textit{R} scenario for $Z$-fermion couplings.
This figure shows that having electron-beam polarization alone already makes a sizable improvement in sensitivity compared with no polarization at all (comparison of the \ILCnopol case and the first column of ILC250 in the plot). Adding positron-beam polarization enhances the sensitivity at low energies. However, at high energies, the positron-beam polarization becomes less critical. As observed in Fig.~\ref{fig:theory1}, the sensitivity depends on the polarization and increases with the \cme. Including the ILC500 program in the ILC250 expectations allows the inspection of all studied GHU models up to the \todo{$B_2^\pm$ models, which predict $m_{KK}=19$ TeV.
Adding also the estimate for ILC1000, the discrimination of $B_3^\pm$ ($m_{KK}=25$ TeV) will also be within reach.}

\begin{figure}[!ht]
    \centering
    \begin{tabular}{c}
            \includegraphics[width=0.45\textwidth]{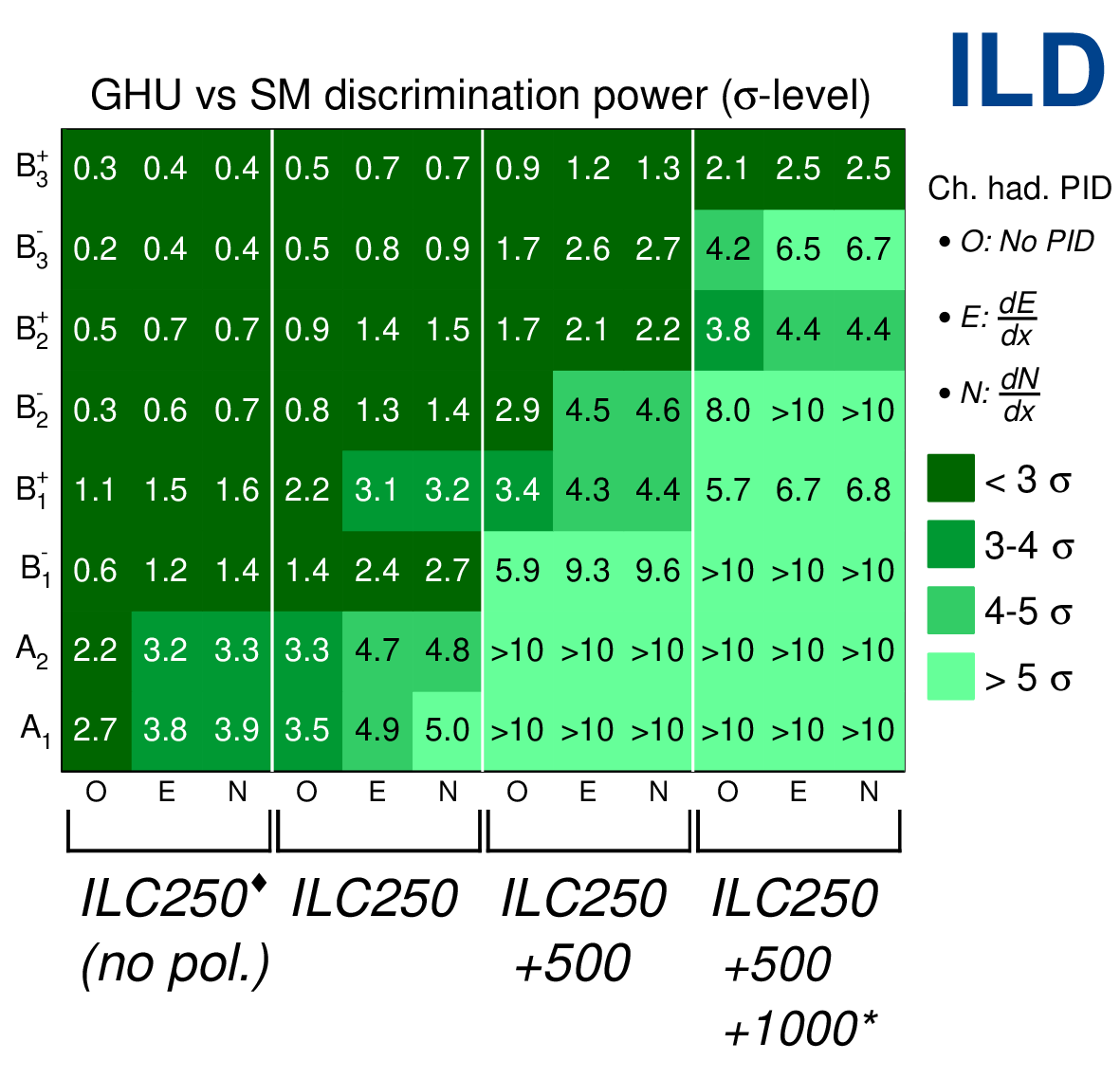} 
    \end{tabular}
    \caption{Similar to Fig.~\ref{fig:SM_vs_GHU} but fixing the \textit{R} case for the $Z$-fermion couplings. Instead, three different scenarios for the charged hadron particle identification capabilities (PID) are considered: \textit{O} for no PID used, \textit{E} for PID based on the ILD baseline \dEdx reconstruction, and \textit{N} for an optimized TPC design with higher granularity and cluster counting reconstruction \dNdx.}
    \label{fig:SM_comparison_PID}
\end{figure} 

\subsection{The role of charged-hadron identification with the ILD TPC}
\label{sec:analysis_ILC250_TPC}

In this study, we have assumed a novel TPC design that will allow cluster counting (\dNdx) for charged-hadron identification capabilities, with better resolution than using the mean energy loss per distance (\dEdx), described in detail in Sect.~\ref{sec:reco}. 
We also consider two other scenarios: a first without any charged-hadron identification (although the TPC tracking capabilities are still used for track reconstruction) and a second one using the traditional \dEdx for particle identification. 
In Ref.~\cite{Irles:2023nee}, it has been estimated that not using charged-hadron identification for the kaon selection could increase the statistical uncertainty by up to a factor of two, especially for the \cquark case, which largely relies on the \Kc, reducing the power of discovery of GHU in these signatures in the baseline program of ILC. 
The expectations for these three cases are shown in Fig.~\ref{fig:SM_comparison_PID}, assuming the $R$ scenario for the $Z$ couplings. The benefits of using charged-hadron identification capabilities with the ILD TPC become clear, especially at the lower energy stage of ILC, since it allows the use of two methods for charge determination, \Bc and \Kc.
The discrimination power differences between using \dEdx and \dNdx are moderate and depend on the different models: the ones predicting larger deviations for the \cquark case are more sensitive to moderate improvements in the \Kc. 

\begin{figure*}[!ht]
    \centering
    \begin{tabular}{cc}
            \includegraphics[width=0.45\textwidth]{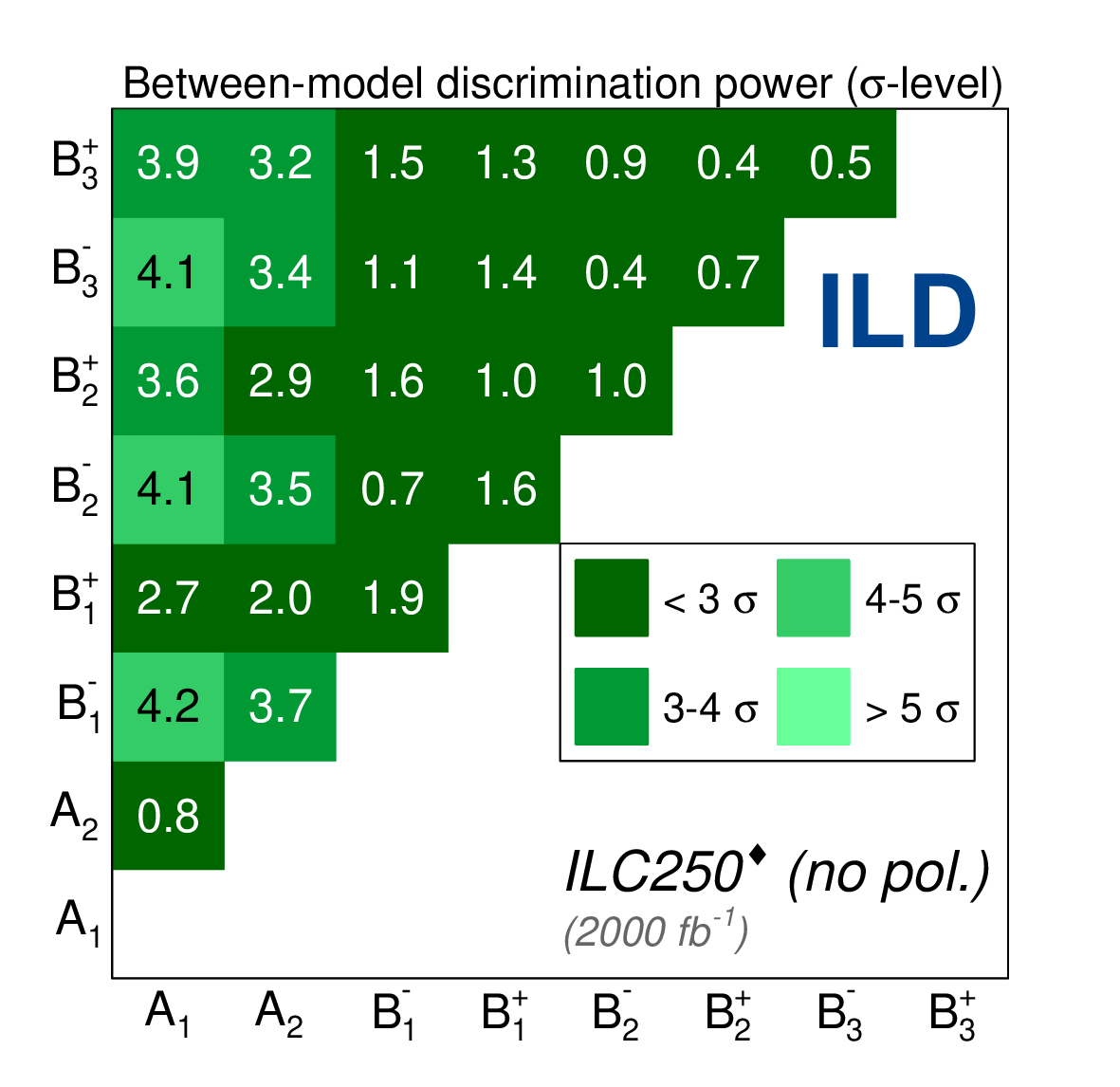}& 
            \includegraphics[width=0.45\textwidth]{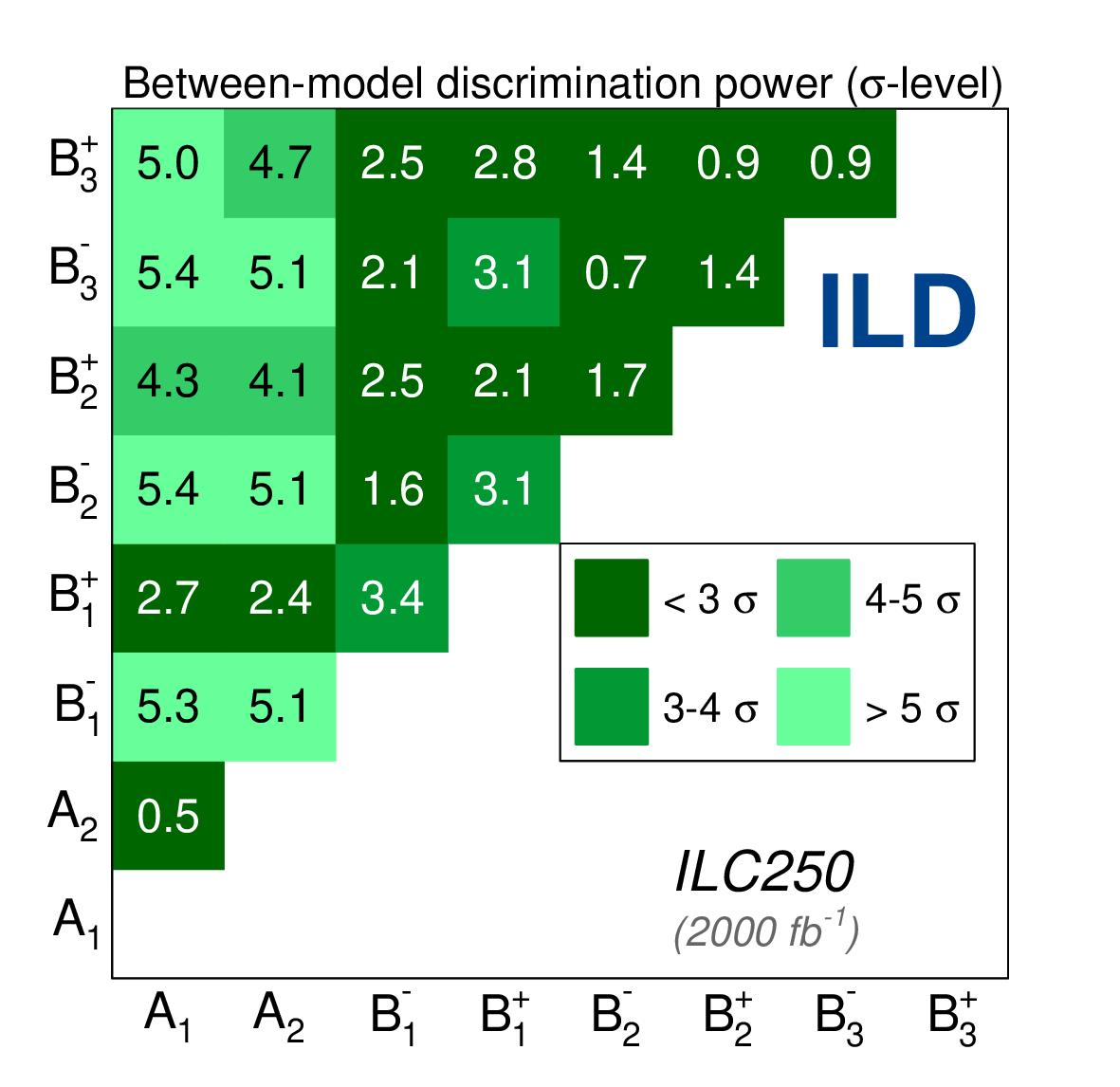} \\
            \includegraphics[width=0.45\textwidth]{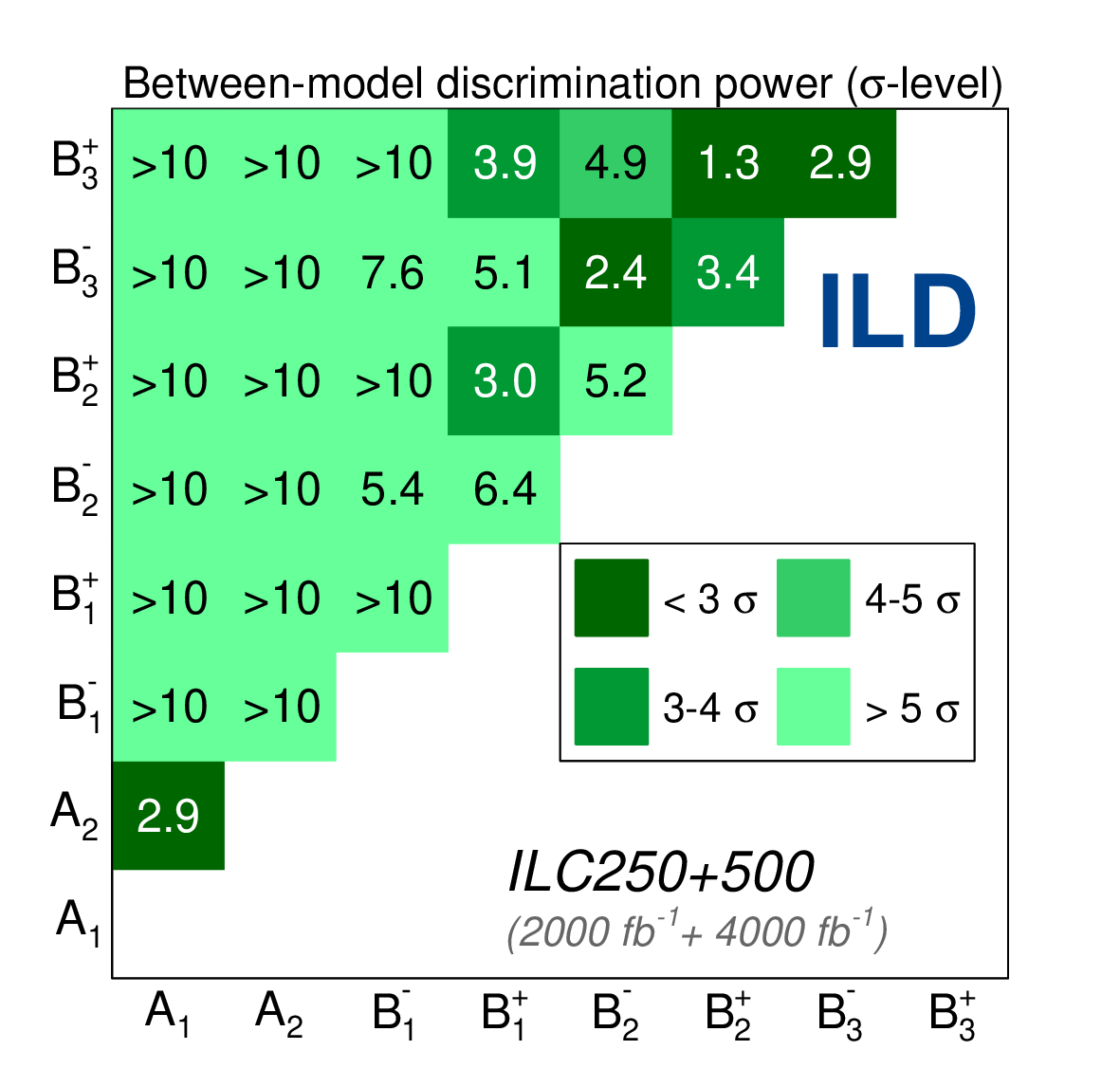} &
            \includegraphics[width=0.45\textwidth]{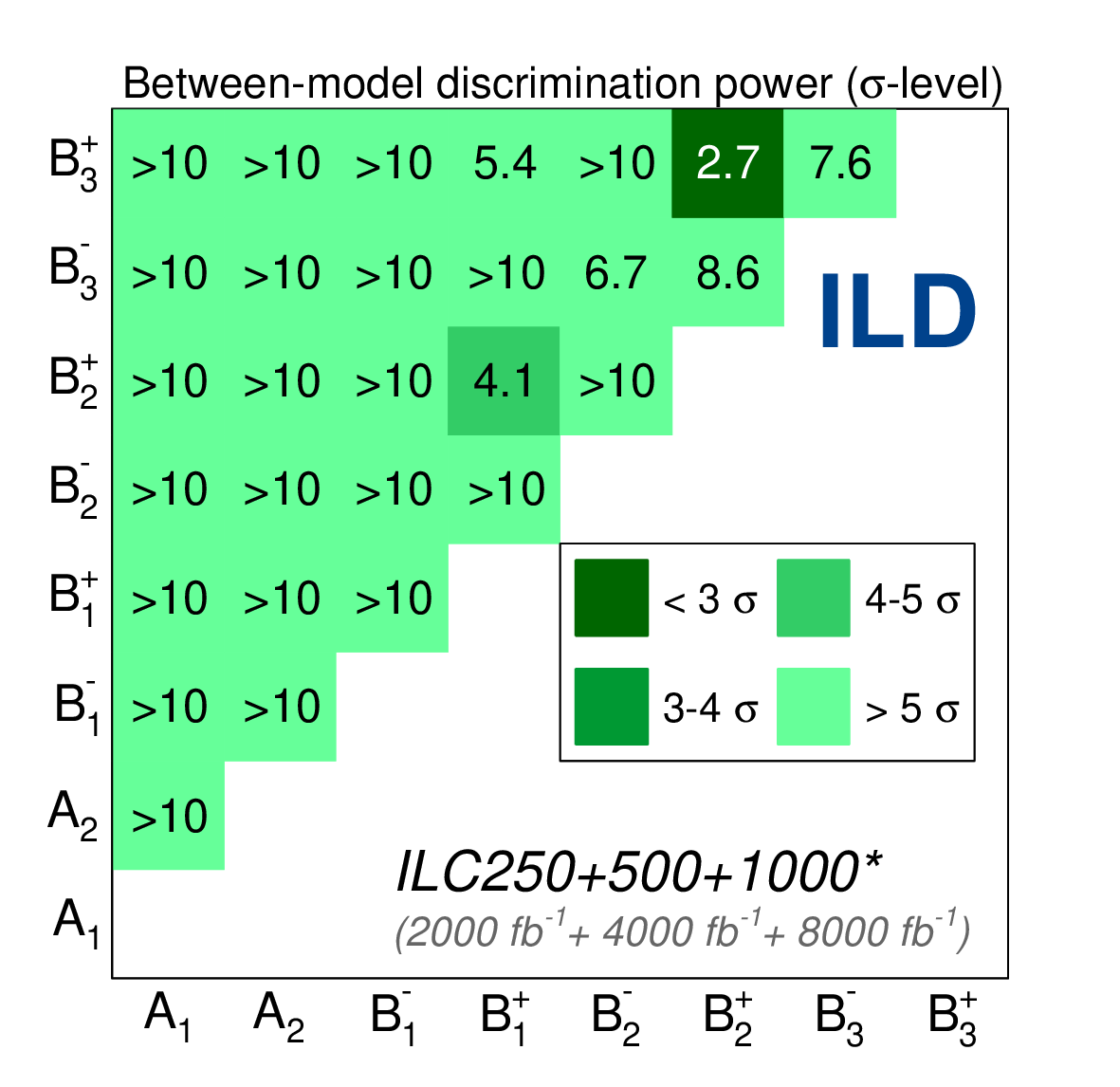}
    \end{tabular}
    \caption{Statistical discrimination power between GHU models after different ILC stages. For completeness, a hypothetical ILC250 stage assuming no longitudinal beam polarization is included. The \ILCnopol, ILC250, and ILC500 estimations are performed using full simulation studies. The ILC1000* is obtained from extrapolations of the ILC500 studies.}
    \label{fig:GHU_discr}
\end{figure*}

\subsection{Discrimination power between GHU models}
\label{sec:analysis_GHU}

In Fig.~\ref{fig:GHU_discr}, the statistical discrimination between the different GHU models is shown for different ILC scenarios. These four plots show the benefits of high longitudinal polarization for both beams and, especially, the benefits of the energy reach foreseen for the ILC.
With operation at 500 GeV and above, almost full differentiation between the different models will be possible, allowing for detailed scrutiny of potential contributions from heavy resonances.

\section{Summary and outlook}
\label{sec:summary}

The search for new physics at the LHC and at future electron-positron colliders requires a global approach.
Searching for new resonances is and will be addressed by a combination of direct searches for such resonances and the precise measurement of observables whose deviations with respect to SM predictions are sensitive to new physics.
The ILC program will provide a broad range of experimental measurements that can be used to probe for BSM physics.
The GHU models discussed here are especially sensitive to deviations in the electroweak observables at high \cme with polarized beams.

This study uses forward-backward asymmetries of $b$ and $c$-quarks in high-energy electron-positron collisions to demonstrate the sensitivity to GHU-inspired models. The experimental input is based on detailed simulations of the ILD at center-of-mass energies of 250 and 500 GeV and extrapolated to 1000 GeV.
More specifically, we present the fully differential cross-section $d\sigma/d \costheta$ from which the \AFBb and \AFBc are inferred.
Studies at ILC250 and ILC500 have been completed, showing that a per-mil level of statistical precision is achievable.
Experimental systematic uncertainties have been briefly discussed. 
These are found to be sub-dominant, thanks to: a) the excellent vertexing and flavor-tagging capabilities expected at the ILC, b) the use of fully differential measurements; 
and c) the use of double-tagging and double-charge measurements that reduce the use of Monte Carlo tools to the minimum when addressing modeling uncertainties, such as the hadronization uncertainties.

Moreover, the ILD also offers the critical capability of providing charged-kaon identification over a broad momentum spectrum.
This enhances the statistical efficiency of these measurements, and, in particular, it provides a factor $\sim2$ improvement for \AFBc.
This analysis uses a simple approach based on selection cuts in kinematic distributions for the background rejection.
Machine learning techniques like multi-classifying neural networks can improve the event selection and construct control regions for the backgrounds~\cite{CMS:2018hnq}. 
This will further improve the sensitivity of the analysis.

The analysis used currently available tools for tracking, vertexing, particle flow, jet reconstruction, and flavor tagging.
In particular, the flavor-tagging algorithm is based on multivariate analysis using boosted decision trees as classifiers~\cite{Suehara:2015ura}.
Advanced machine-learning methods such as graph neural networks are expected to significantly improve jet flavor identification performance~\cite{ATLAS:2022rkn}.

The GHU models described in Refs.~\cite{Funatsu:2017nfm,Funatsu:2020haj,Funatsu:2023jng} show high expected sensitivity for the \AFBb and \AFBc observables.
The expected sensitivity increases with the energy and depends on the electron and positron beam polarization.
These GHU models predict new massive $Z^\prime$ resonances and deviations of all SM $Z$-fermion couplings.
They are constructed not to induce significant changes in the EW precision observables measured at past lepton colliders and agree with the non-observation of $Z^\prime$ at LHC.

We show that the ILC operating polarized beams colliding at 250 GeV and 500 GeV could provide full discrimination power (at the $5\sigma$ level) between these models and the SM, through \AFBb and \AFBc measurements. 
The ILC250 case has also been compared with an ILC250 without beam polarization. For the latter case, at least a factor of two of integrated luminosity is required to get similar prospects.

The ILC program also considers a run at high energies of 1000 GeV. 
Detailed studies of these scenarios, with optimized designs of the ILD,
would be required to provide realistic estimations of the uncertainties expected in these cases. 
Instead, the exercise of extrapolating from the ILC500 full simulation studies has been performed.
Including the predictions for ILC1000, full discovery potential for different GHU models predicting Kaluza-Klein resonance with masses up to 25 TeV will be possible.

It is important to remark on the intrinsic importance of precisely constraining the couplings of the $Z$-boson to all fermions without deviations caused by heavy resonances predicted by BSM. 
This can be done at dedicated runs at the \Zpole (ILCGigaZ) or with a dedicated analysis at ILC250 using radiative return events~\cite{ILCInternationalDevelopmentTeam:2022izu}. 
The importance of getting updated estimations for such couplings with at least one order of magnitude better precision than current measurements is also shown in this study.

This study is based on leading-order accurate simulations, and calculations should be updated, including next-to-leading order contributions in QCD and EW at the theory and Monte Carlo simulation level. 
This is out of the scope of this work due to the non-existence of such simulations or GHU predictions at the moment.

Measurements for different fermions, various types of observables, under different \cme and beam-polarization conditions will allow a deep investigation of such theories and the disentanglement of potential new effects: deviations of the $Z$-$f$ couplings, contributions from new heavier resonances, mixing effects, etc. 
In particular, future colliders will provide multiple differential cross-section measurements for different processes that can be used to set limits on the SM via effective-field interpretation models.
In the Standard Model Effective Field Theory (SMEFT)~\cite{Brivio:2017vri}, the study of the implications of new physics is performed indirectly through the introduction of modifications to the SM at very high energies.
The strength of the modification of SM couplings is determined by the energy scale at which the new physics is expected and the coupling strengths expressed by the Wilson coefficients, which can be fitted, or the exclusion limits on the SMEFT parameters and, therefore, BSM physics can be set. 
The SMEFT approach has the advantage of allowing the simulation of the impact of the modification of couplings for several physics processes simultaneously. Therefore, a global analysis of the data with different processes is possible, accounting for their correlations.




\begin{acknowledgements}

We are grateful to the ILD Publication and Speakers Bureau, particularly K. Kawagoe, M. Berggren, D. Jeans
and K. Fuji for their work and support as the editorial board team. We thank Y. Hosotani for his indispensable input.
We thank the LCC generator working group and the ILD software working group for providing the simulation and reconstruction tools and producing the Monte Carlo samples used in this study.
This work has benefited from computing services provided by the ILC Virtual Organization, supported by the national resource providers of the EGI Federation and the Open Science GRID.
A.I. and J.P.M acknowledge the support by the PlanGenT program from the Generalitat Valenciana (Spain) with the grant number CIDEGENT$/2020/021$.
A.I. acknowledges the MCIN with funding from the European Union NextGenerationEU and Generalitat Valenciana in the call Programa de Planes Complementarios de I+D+i (PRTR 2022) (project \textit{Si4HTopF} with reference ASFAE$/2022/015$). A.S. acknowledges support from the Alexander von Humboldt Foundation through the Feodor Lynen Research Fellowship. A.I, J.P.M and A.S acknowledge the funding from  PID2021-122134NB-C21. N.Y. acknowledges the support by the Ministry of Science and Technology of Taiwan under Grant No. MOST-111-2811-M-002-047-MY2 and the Japan Society for the Promotion of Science (JSPS) KAKENHI Grant No. 21H05182.

\end{acknowledgements}





\end{document}